\documentclass[manuscript,screen]{acmart}
 
\AtBeginDocument{%
  \providecommand\BibTeX{{%
    \normalfont B\kern-0.5em{\scshape i\kern-0.25em b}\kern-0.8em\TeX}}}

\usepackage{tabularx}

\usepackage{pgfplots}

\usepackage{tikz,pgfplots,pgfplotstable}

\usepackage{multirow}
\usepackage{graphicx}
\usepackage{pifont}
\usepackage{subcaption}
\usepackage{cleveref}
\usepackage{rotating}

\usepackage{makecell}

\begin{document}

\title[A Survey of AI Reliance]{A Survey of AI Reliance}

\author{Sven Eckhardt}
\email{eckhardt@ifi.uzh.ch}
\orcid{0000-0002-4713-8408}
\affiliation{%
  \institution{University of Zurich}
  \city{Zurich}
  \country{Switzerland}
}
\author{Niklas K\"uhl}
\email{kuehl@uni-bayreuth.de}
\orcid{0000-0001-6750-0876}
\affiliation{%
  \institution{University of Bayreuth \& Fraunhofer FIT}
  \city{Bayreuth}
  \country{Germany}
}
\author{Mateusz Dolata}
\email{dolata@ifi.uzh.ch}
\orcid{0000-0002-2732-4465}
\affiliation{%
  \institution{Zeppelin University}
  \city{Friedrichshafen}
  \country{Germany}
}
\affiliation{%
  \institution{University of Zurich}
  \city{Zurich}
  \country{Switzerland}
}
\author{Gerhard Schwabe}
\email{schwabe@ifi.uzh.ch}
\orcid{0000-0002-0453-9762}
\affiliation{%
  \institution{University of Zurich}
  \city{Zurich}
  \country{Switzerland}
}

\begin{abstract}
Although artificial intelligence (AI) systems are becoming increasingly indispensable, research into how humans rely on these systems (\textit{AI reliance}) is lagging behind. To advance this research, this survey presents a novel, comprehensive sociotechnical perspective on AI reliance, essential to fully understand the phenomenon. 
To address these challenges, the survey introduces a categorization framework resulting in a morphological box, which guides rigorous AI reliance research.
Further, the survey identifies the core influences on AI reliance within the components of a sociotechnical system and discusses current limitations alongside emerging future research avenues to form a research agenda. 
\end{abstract}




\maketitle

\section{Introduction}

The term \textit{Artificial intelligence (AI)} was first proposed at the Dartmouth Conference in 1956 \cite{mccarthy2006proposal}. Since then, considerable progress has been made toward the development of modern AI systems. However, the full implications of human \textit{behavior} with regard to these novel AI systems remain poorly understood. It is possible that individuals may use these systems to improve their performance or as lazy shortcuts, as evidenced by reduced cognitive effort when confronted with AI advice \cite{gajos2022people}. Therefore, not only is the performance of AI systems relevant, but also the reliance of users serving as decision-makers on AI advice---a phenomenon we call \textit{AI reliance}. AI systems do not exist in isolation, but rather, they interact with humans. Together with the AI system, humans form a sociotechnical system (STS). Therefore, when interested in AI reliance, one should not only consider the technical capabilities of AI systems or the human properties alone. However, the STS includes more than just the human user and the AI system: it also includes the reliance interaction and the general environment in which AI reliance occurs. To get a comprehensive perspective on AI reliance, these four components need to be considered at once.

The pervasive use of AI in research and practice has led to a growing interest in the topic of AI reliance in order to understand how users behave when confronted with AI systems and their recommendations. Based on early conceptualizations of what is essentially AI reliance \cite[e.g.,][]{lee2004trust,dzindolet2003role}, we present key terminology of AI reliance. If a user follows the AI advice, we can infer that the user \textit{relies} on that advice. Conversely, if a user follows incorrect advice, we can infer that the user is \textit{overrelying} on the AI advice. If a user does not follow correct advice, we can infer that the user is \textit{underrelying} on the AI advice. Only by taking into account AI reliance can we investigate whether people are using the systems appropriately, blindly following them, or potentially ignoring their recommendations entirely. 

An illustrative example of the role of AI reliance is the case of the COMPAS \cite{brennan2018correctional} system, an AI system designed to classify the recidivism risk of criminal defendants. In this example, overreliance and underreliance have a significant impact on humans who may be sentenced to longer prison terms. This example highlights the core challenge of AI reliance in practice: achieving \textit{appropriate reliance}. The human-AI team can then achieve a complementary performance that surpasses that of either the human or the AI alone \cite{bansal2021does}. This increases the overall performance of the human-AI collaboration compared to that of each individual component. 

Despite the overall importance of AI reliance, there is no clear guidance on how to conduct AI reliance research. For instance, researchers might face multiple inconsistent definitions and measures, of which some remain niche. There are several theoretical and conceptual considerations around AI reliance \cite[e.g.][]{solberg2022conceptual,kerasidou2022before,rao2022reliance}. These considerations mostly aim at differentiating trust and reliance and base their conclusions on the philosophical literature around reliance in a more general sense. A clear focus on \textit{AI} reliance is missing. Whereas some recent approaches aim to formulate a \textit{formal definition of reliance} \cite{guo2024decision}, they focus solely on the definition of AI reliance and look at how reliance has been established in past research. Consequently, while they offer a theoretically sound conceptualization, this might not reflect how AI reliance has been understood in research. 

Accordingly, researchers who want to engage with this topic might find themselves torn between theoretical perspectives and their peers' research practice. Researchers who are studying AI reliance are still navigating uncharted waters in their efforts to identify the most appropriate methodology and fruitful avenues for conducting AI reliance research. Overall, the current state of AI research can be described as rather chaotic and unstructured, with a variety of definitions, measures, conceptualizations, and general understandings. In the long run, this may result in the replication of unsound practices or the application of research methods without an understanding of their implications. A systematic understanding of researching AI reliance is therefore important, and is presented in this review by using the following research questions:
\begin{itemize}
    \item[\textbf{RQ1}] \textit{What is AI reliance according to current scientific discourse?}
    \item[\textbf{RQ2}] \textit{Which components of a sociotechnical system have been identified as primary factors influencing AI reliance?}
    \item[\textbf{RQ3}] \textit{How can rigorous research on AI reliance be effectively conducted?}
    \item[\textbf{RQ4}] \textit{What is a research agenda for AI reliance?}
\end{itemize}
We address these research questions by conducting a systematic review of existing literature on AI reliance. First, the survey provides an overview of what AI reliance is according to current scientific discourse (\textbf{RQ1}). In doing so, the survey introduces a sociotechnical perspective on AI reliance, which is essential for a thorough understanding of this topic, and presents components that are currently identified as primary factors influencing AI reliance (\textbf{RQ2}). Second, the survey presents a categorization framework and a morphological box that researchers can use as a template for conducting well-founded AI reliance studies. (\textbf{RQ3}). Finally, the survey discusses the limitations of the current literature and future research in the form of a research agenda (\textbf{RQ4}).

This survey supports two distinct groups of AI reliance researchers. The first group is interested in extending the field. For this group, the survey presents a research agenda for conducting research on this topic. The second group is interested in fostering appropriate reliance on their AI systems. For this group, the survey provides an overview of current approaches and a framework and morphological box to use when conducting AI reliance research. 
 
\section{Background on AI Reliance}
\label{sec:related_work}
\subsection{Related Literature Surveys}

The field of AI is missing unified and widely accepted conceptualizations and definitions. However, research in this field is closely connected to other fields within human-centered AI. Here, we provide a brief review of the literature on related concepts and their links to AI reliance. A summary of these literature surveys in tabular form is shown in the digital appendix. Once modern AI systems achieved sufficient task performance to be used in productive settings, practitioners and researchers began to focus on the human-interaction aspects of these systems. Most research in this field has focused on the explainability of AI systems to influence the human use of these systems. One of the earliest reviews in this area was conducted by \citet{adadi2018peekin} in 2018, who were among the first to systematically structure the existing literature on explainable AI. More recently, \citet{rudresh2023explainable} published a comprehensive overview of the core ideas, techniques, and solutions associated with explainable AI. Another relevant topic is the fairness of AI systems, which also might incline humans to use the system more. In general, it is desirable that AI systems output fair decisions that do not discriminate. There is, however, no single definition of fairness but rather several concepts of fairness, which, for example, \citet{mehrabi2021survey} or \citet{pessach2022fairness} surveyed. While these concepts might influence reliance, the link to AI reliance remains underexplored. 

Explanations, fairness, and related concepts are often employed to achieve trustworthy AI systems, as surveyed in \citet{kaur2022trustworthy}. They present one of the first sociotechnical considerations of the interplay between user and AI system, without explicitly stating the perspective. While the aforementioned studies focus on the overarching theme of trust, this review will concentrate on reliance as behavior. With regard to human-AI decision-making, the survey by \citet{lai2023towards} represents one of the first attempts to examine the existing literature in a structured manner. The behavior of humans is also examined more closely there. While this behavior is relevant to the present study, the study lacks a clear focus on reliance and only considers literature up until 2021. 

In general, research on fair, responsible, and explainable AI has shown clearly that we need to frame complex phenomena related to AI as contextualized in the usage practice and organizational context, which may be achieved by the STS lens \cite{kudina2024sociotechnical,dolata2022sociotechnical, vassilakopoulou2022responsible}. However, existing conceptual work and overviews of AI reliance focused on technical or theoretical definitions \cite[e.g.][]{guo2024decision}. Therefore, we follow the sociotechnical perspective to offer a comprehensive, overarching view of AI reliance, comprising studies from various subfields. 

\subsection{Background on AI Reliance}
The notion of reliance is not unique to AI or computer systems---it is extensively discussed in the field of philosophy and psychology. Reliance is often aligned with trust, as the two concepts appear similar at first glance \cite{hawley2014trust}. However, many philosophers argue that trust cannot be expressed toward inanimate objects \cite[e.g.][]{hawley2014trust, holton1994deciding}. Nevertheless, a substantial body of AI research has been conducted on trust in AI, often referred to as trustworthy AI \cite{kaur2022trustworthy}. It can be argued that computer systems (and AI) are not inanimate objects, as evidenced by the ``computers are social actors'' principle \cite{nass1994computers}. When computer systems research discusses trust, it is, however, often concerned with reliance \cite{deley2020assessing}. In this article, we follow the line of discourse that suggests differentiating between trust (an emotional or attitudinal stance toward something) and reliance (observable behavior) \cite[e.g.][]{lee2004trust}. This distinction is particularly important, as a person may have trust without reliance (or vice versa). Revisiting the COMPAS example, a judge might distrust the system due to a bad past experience. However, the judge may still rely on it and base their sentencing on it for various reasons, such as a lack of alternative or delegation of responsibility. This highlights that attitudes and behavior are not necessarily aligned. Therefore, this review focuses on studies that investigate behavior rather than attitude.

The discussion around reliance is not unique to the topic of AI. Therefore, even before AI systems, researchers discussed the topic of reliance on computer systems automating tasks, so-called automation systems \cite{sheridan2002humans}. Automation is defined as \textit{``the execution by a machine agent (usually a computer) of a function that was previously carried out by a human''} \cite[][p. 231]{parasuraman1997humans} \footnote{This notion of automation systems is evolving over time with the capabilities of computer systems. In this survey, we differentiate automation systems from AI by their focus on automating tasks and lack of predictive power. We also acknowledge that this distinction is becoming increasingly blurred.} A simplified example of an automation system where the topic of reliance becomes important is a conveyor belt with a warning light that turns on once there is overheating. In that example, reliance can be defined as the human intervening as soon as the light turns on. However, previous research has frequently concentrated on the concept of \textit{trust} in human-machine interaction in automation systems \cite[e.g.][]{muir1994trust,muir1987trust}. A frequently researched case is that of pilot cockpit automation systems \cite[e.g.][]{cafarelli1998effect, wickens1995designing}. When reading these earlier studies, it becomes evident that the concept of trust is more complex than previously thought. For example, \citet[][p. 366]{wickens1995designing} defines trust as \textit{``the extent to which the pilot believes that, and \textbf{behaves} as if the automation will carry out its assigned task in a reliable fashion.''} This definition encompasses not only the pilot's attitude toward the system but also their behavior. 

An early definition of what is essential reliance on automation systems has been considered by \citet{parasuraman1997humans}, by distinguishing between attitude and \textit{use}. They define the \textit{use}, \textit{misuse}, \textit{disuse}, and \textit{abuse} of automation systems. They define \textit{use} as \textit{``the voluntary activation or disengagement of automation by human operators,''} misuse as the \textit{``overreliance on automation,''} disuse as the \textit{``neglect or underutilization of automation,''} and abuse as the \textit{``automation of functions by designers and implementation by managers without due regard for the consequences for human performance''} that might also resolve in the misuse and disuse of the system \cite[][p. 230]{parasuraman1997humans}. 

The increasing power of computer systems has led to a shift from deterministic automation to the development of AI systems, which is now most often enabled by machine learning methods \cite{goodfellow2016deep}. AI use gives these systems the ability to predict outcomes and distinguishes them from automation systems. In the case of the COMPAS algorithm \cite{brennan2018correctional}, the primary objective was not the automation of sentencing, but rather to predict whether an offender would reoffend after being released. This AI system provided advice to a human judge, who made the decision after being presented with the AI advice. Some research about AI reliance has followed the early notions of reliance on automation systems and defined reliance as \textit{the following of AI advice} \cite{lee2004trust}. This reliance, however, does not indicate whether this was the correct decision or not. It is often the case that AI systems are probabilistic in nature and can therefore provide erroneous advice. Consequently, the concepts of underreliance and overreliance become important. Overreliance (\textit{misuse} of AI advice) refers to the acceptance of incorrect advice. Conversely, underreliance (\textit{disuse} of AI advice) refers to the rejection of correct advice. In a manner analogous to the definitions of automation systems, appropriate reliance can be defined as the acceptance of correct AI advice and the rejection of incorrect AI advice. This understanding of reliance can be measured and quantified by counting the number of times a user follows the AI advice. In the literature on AI reliance, this is often referred to as \textit{the agreement fraction} or \textit{agreement percentage}. \cite[e.g.][]{vasconcelos_explanations_2023, lai_selective_2023}. With this understanding, it is possible to measure underreliance (the number of instances where the user does not follow the AI's correct advice), overreliance (the number of instances where the user follows the AI advice despite it being incorrect), and appropriate reliance (the number of instances where the user follows the AI's correct advice and does not follow incorrect advice). 

However, presenting advice may not be exactly equated with automation systems. Therefore, a more recent angle of reliance on AI systems is to lend insights from advice-taking literature. In advice-taking literature, the concept of the judge-advisor system (JAS) is widespread \cite{sniezek1995cueing}. In a JAS, the judge gets a task and receives advice from an advisor. Based on this task and advice, the judge makes the final decision. This JAS also introduces a new angle on advice-taking, where the judge gives an initial estimate without being exposed to the advice. Then an advisor is employed to advise the judge on their decision, whereafter the judge can give a final decision. In the case of the COMPAS algorithm \cite{brennan2018correctional}, the judge would first have to make their own assessment. Only then would the advice of the COMPAS system be presented to the judge, who then makes the final decision. 

This JAS can be transferred to AI advice, where the AI system serves as an advisor to the human (judge) by providing advice. This can provide a more granular view of AI reliance. Some research suggests that reliance can only be observed if the initial decision by the human differs from the AI advice \cite[e.g.,][]{schemmer_appropriate_2023}. If human and AI decisions coincide, it is unclear whether the human relied on the AI or if it was coincidental. \textit{Overreliance} is defined when a human changes from a correct to an incorrect decision after incorrect AI advice. \textit{Underreliance} is defined when a human fails to change from an incorrect to a correct decision despite a correct AI advice. Finally, \textit{appropriate reliance} is defined when the human does not change to an incorrect decision after incorrect AI advice, or change to the correct decision after correct AI advice. This conceptualization of an initial human decision permits a deeper look into AI reliance. Rather than counting instances of agreement with AI advice, some researchers are interested in the number of instances where the users change their decision after receiving AI advice. This phenomenon is often referred to as the \textit{switch fraction} or the \textit{switch percentage} \cite[e.g.][]{zhang_you_2022, schmitt_towards_2021}. Moreover, reliance can be quantified in the absence of discrete decision-making or even regression cases. This is derived from the literature on advice-taking, whereby the \textit{weight of advice (WOA)} is calculated by: \textit{(final estimate - initial estimate) / (advice - initial estimate)} \cite{sniezek1995cueing}. 

These insights highlight major disagreements within AI reliance research. Depending on the perspective on AI reliance, research methodologies and subsequent findings may differ. This highlights the complexity of AI reliance research and the need to consider AI reliance from multiple perspectives. AI reliance is inherently sociotechnical, requiring a multifold perspective.. Focusing on one aspect of the STS is insufficient. For instance, the design of an AI system can produce different results based on the human users, the interaction between the AI and the humans, and the environment. Similarly, a human user may exhibit different levels of AI reliance depending on the AI system's design, how they interact with it, and their surroundings. The same holds true when changing the interaction itself or the surrounding environment. In the following section, we will introduce the perspective of STS, its four major components, and their relation to AI reliance.

\subsection{Sociotechnical Systems}

An STS consist of a \textit{technical component} (e.g., the AI system), a \textit{social component} (e.g., the human decision-maker), and an \textit{interaction} between these two in a broader context or \textit{environment} (e.g., an organization). When considering AI reliance, it is important to recognize that the final result of the interaction between the AI system and the user occurs in a specific environment and is contingent upon both the human decision and the AI advice. A sociotechnical lens is therefore essential. The individual components are introduced in the following section and used as a basis for the analysis of the literature on AI reliance. The technical component is defined as comprising technical and material artifacts and the techniques or practices employed to use the artifact \cite{trist1951some,lee2004thinking}. In the case of human-AI interactions, this would primarily be the design and advice type of the AI system. The social component encompasses individuals or collectives, as well as the relationships between them, which may be expressed as roles, hierarchies, structures, economic systems, cultures, power relations, or communication networks \cite{leonardi2012materiality, sarker2019sociotechnical}.  In the context of human-AI interaction, this would be the human user/decision-maker. The STS is, however, more than the sum of the technical and social components. These components enter complex mutual interactions, therefore, an understanding of the system's functioning is only possible if one considers the interdependencies between the components and their subcomponents \cite{leonardi2012materiality}. An analysis that focuses on only one of the components offers an incomplete perspective of the system, which might limit insights for identifying, for example, the overall system outcomes. The interaction between the social and technical components is recursive and reciprocal, forming a process of joint optimization \cite{sarker2019sociotechnical}. An STS also operates in a broader societal, political, regulatory, organizational, and historical context \cite{chatterjee2021possible}. It is essential to consider the interaction with the environment, which can be framed in terms of an input-transformation-output model. Consequently, an STS receives inputs from the environment, such as economic pressure or higher demand, which it transforms to deliver the output that fits the environment. 

\subsection{Summary}

The current research on AI reliance aims to understand and design for appropriate reliance on AI advice \cite[e.g.][]{benda2022trust}. The goal of this research is, typically, to achieve complementary team performance, where the team of humans and AI exceeds the performance of the human or AI components if considered separately \cite{bansal_does_2021}. There is no consistent approach across the related work. While there have been reviews analyzing human-AI decision-making \cite[e.g.][]{lai2023towards}, there has been no focus on the reliance of decision-makers on AI advice. These studies frequently focused on a single aspect of the STS, namely the technical component, i.e., the AI system itself. The shortcoming of this approach is exemplified by the case of explanations \cite[e.g.][]{schemmer_appropriate_2023, bucinca_trust_2021} that often show inconclusive results \cite{schemmer2022meta}. We argue that one reason is that a comprehensive perspective is missing. The STS perspective enables us to introduce a comprehensive perspective and analyze existing literature on AI reliance structurally. This allows us to establish uniform approaches and provide guidance, essential for research to have a real-life impact and achieve the desired appropriate reliance on AI advice. 

\section{Survey Methodology}
\label{sec:method}

Given the lack of unified approaches in the literature, a structured literature review is employed to provide an overview of the current approaches that researchers who investigate AI reliance employ. This section describes the methodology used for structurally querying the literature. Based on a classification framework introduced in  \Cref{sec:classification}, the results of the structural literature review are presented in \Cref{sec:results}. These results are then used in \Cref{sec:critical_review} to discuss the shortcomings and limitations of current approaches and to present guidance for researchers conducting sound AI reliance research. Additionally, future avenues for further developing the field of AI reliance are presented.

\subsection{Database Queries}

\subsubsection*{Keywords}

To identify relevant articles for research into AI reliance, we employed a structured query of scientific databases. As there are numerous underlying technologies for AI systems, we also queried the common terms that are treated as synonyms, such as machine learning and deep learning\footnote{While we are aware of the technical distinction between artificial intelligence and machine learning as discussed, for example by \citep{kuhl2022artificial}, we acknowledge that the synonyms have established themselves in the computer science and human-computer interaction communities.}. In order to include articles that explicitly investigate overreliance and underreliance, we added these terms to the search string:  \texttt{{(}Artificial Intelligence {OR} AI {OR} Machine Learning {OR} ML {OR} Deep Learning {OR} DL{)} {AND} {(}Reliance {OR} Overreliance {OR} Underreliance{)}}. The instantiated search strings for the databases are presented in the appendix. 

The search was conducted on titles, abstracts, and author keywords. Hyphenated compound words, such as ``over-reliance'' are also included with the given search string, negating the necessity to include these words explicitly in the search string. Including the terms ``rely'' and ``relies,'' would, however, lead to many results unrelated to AI reliance, which is why we therefore opted to only include reliance to focus on articles that explicitly investigate that concept. Further, we focus solely on the term ``reliance'' and its related forms, despite the existence of closely related terms such as ``(objective) trust'' or ``trusting behavior''. This is because our goal is to understand how the concept of reliance is defined and used in current research. Researchers who use different terms often do so to highlight a different perspective, i.e., focusing on intention, attitude, or opinion. Including these terms could therefore distort the specific focus of our review.

Four scientific databases in the field of computer science and human-computer interaction were queried, with the cutoff date for the selected article set to 31 December 2023. The databases include the \textit{ACM Digital Library}, \textit{AISeL}, \textit{IEEE Xplore}, and \textit{Scopus}. The rationale is that the first three collectively represent the most important outlets in the categories of computer science, human-computer interaction, and information systems, while SCOPUS provides a broader view of existing literature\footnote{Given the vast scale of SCOPUS, we focused only on the top 25\% of outlets according to their citation score in the categories of \textit{computer science}, \textit{business}, \textit{management and accounting}, \textit{decision sciences}, \textit{multidisciplinary}, \textit{psychology}, \textit{general social sciences}, and \textit{human factors and ergonomics}.} Articles that were not initially identified due to this focus will be identified in the subsequent forward and backward searches (\Cref{sec:fw_bw}).

\subsubsection*{Inclusion of Articles}
\label{sec:criteria}

The query yielded 1,337 articles, but only articles concerned with human reliance on AI systems are considered, which limits the number of relevant articles.  This excluded many articles in which the term 'reliance' was mostly used to refer to AI's reliance on datasets, labeled data, or similar resources. Further, this survey focuses on articles researching AI reliance through the lens of human behavior, while articles of a purely conceptual nature are included in the related work section. Finally, to provide a well-grounded overview of modern AI research, we only included peer-reviewed articles in the English language that were published after 2010. We chose 2010 as the cut-off date because this was when the first deep learning models became popular (e.g., illustrated by the advances of the \textit{ImageNet} challenge \cite{deng2009imagenet}), marking a new era in AI research. It is also worth noting that the oldest article identified in this survey was published in 2018, which highlights the fact that the 2010 cut-off date is fitting.

All queried articles were filtered by one author. For most articles, inclusion or exclusion was trivial (i.e., they were not concerned with human reliance on AI systems). Articles for which inclusion or exclusion was not trivial were brought to an author workshop after the first phase, where three authors discussed their inclusion or exclusion. This yielded a total of 45 relevant articles. To ensure that no relevant articles were missed due to narrow keywords or database selection, we conducted forward and backward searches.

\subsubsection*{Forward and Backward Search}
\label{sec:fw_bw}

\begin{figure}[htbp]
    \centering
    \includegraphics[width=0.8\textwidth]{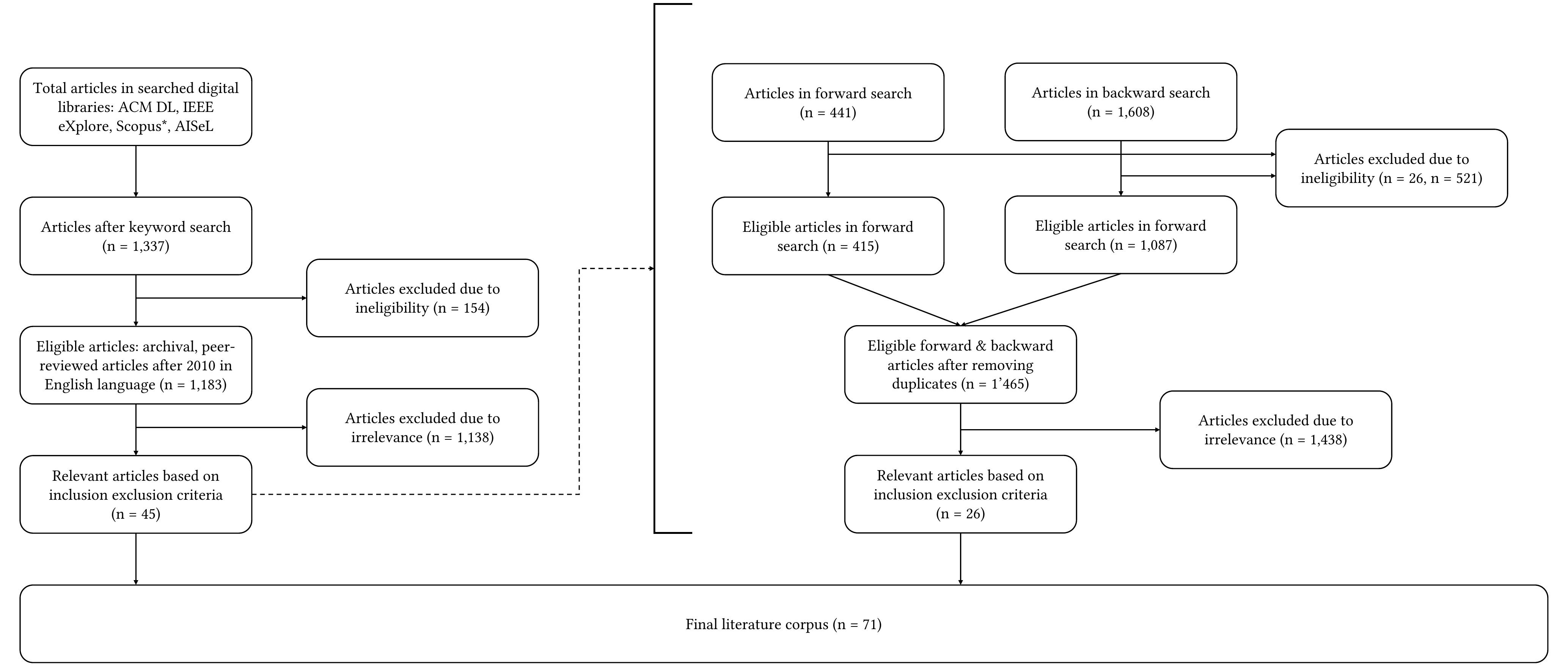}
    \caption{PRISMA flow chart \cite{moher2009preferred} of queried literature}
    \Description{This figure illustrates the PRISMA flow chart for the literature search. It delineates the steps undertaken in the initial database search and subsequent forward and backward searches, which collectively yielded the final set of relevant articles.}
    \label{fig:PRISMA}
\end{figure}

A forward and backward search was conducted after selecting relevant articles based on querying the literature databases. These searches were performed on SCOPUS without any additional filters\footnote{Three articles lacked DOI, preventing their inclusion in the SCOPUS database. To identify additional relevant articles, a forward and backward search was conducted in Google Scholar for the three articles without DOIs. This yielded no additional relevant articles.}. For the forward search, all articles citing the relevant articles were queried, which yielded a list of 441 articles. For the backward search, all articles cited by the relevant articles are included. This yielded a list of 1,160 articles. We merged these two lists, removed duplicate entries, and obtained a list of 1465 articles. The whole process is illustrated in \cref{fig:PRISMA}. After using the same inclusion criteria introduced above, we identified another 26 relevant articles. Two main reasons account for the absence of these articles in the initial search. First, several articles were published in an outlet not included in the top 25\% of SCOPUS. Second, some of these articles did not mention reliance in the title, abstract, or keywords and could therefore not be queried using the above-introduced keywords. It was only in the main body that the human reliance on AI systems was discussed.

\subsection{Query Results}
\label{sec:query_results}

\subsubsection*{Summary of Query} In the following section, we summarize the query results. A total of 71 relevant articles were identified through the application of the described methodologies. The query results are summarized in \Cref{fig:PRISMA} using a PRISMA flow chart \cite{moher2009preferred}. The initial database search yielded 1,337 articles, of which 154 were excluded due to them not being in English, having been published before 2010, or not having been peer-reviewed, leaving 1,183 articles. After screening the articles using the inclusion and exclusion criteria, we identified 45 relevant articles that investigated the empirical use of AI. We then conducted a forward and backward search, which yielded 441 unique articles in the forward search and 1,608 unique articles in the backward search. After excluding ineligible articles, we were left with 415 articles for the forward search and 1,087 articles for the backward search. As some articles were included in both the forward and backward search sets, merging the two yielded 1,465 unique articles. After screening these articles concerning the inclusion and exclusion criteria, we were left with 26 additional articles that concerned the empirical investigation of AI reliance. This resulted in 71 unique articles dealing with the empirical study of AI reliance. The results are summarized in \Cref{fig:PRISMA}. The examination of basic bibliometrics, such as publication year or outlet, provides insight into the communities concerned with AI reliance. This is described in the following section. 

\subsubsection*{Bibliometrics}
\label{sec:bibliometrics}

From 2018 to 2023, the number of published articles increased from one to 31. The yearly counts were as follows: one in 2018, five in 2019, two in 2020, 14 in 2021, 18 in 2022, and 31 in 2023. Further, despite our search including articles from 2010 onwards, the earliest articles identified are from 2018\footnote{The article of \citet{dietvorst_overcoming_2018} was accepted for publication in 2016 but only published in 2018.}. It is evident that the number of published articles has increased significantly since 2021. This suggests that research on AI reliance commenced at around this time. The EU AI Act was also first discussed in 2021 \cite{EU_proposal_2021}, demonstrating the timeliness and relevance of the topic of AI reliance. Further, we find that the most relevant articles originate from HCI literature, which is listed in the ACM Digital Library. The most notable were the  \textit{Proceedings of the ACM on Human-Computer Interaction} with 16 articles; \textit{Proceedings of the CHI Conference on Human Factors in Computing Systems} with eight articles; \textit{Proceedings of the International Conference on Intelligent User Interfaces} with six articles; and ICPS \textit{ACM International Conference Proceeding Series}, with four articles. Furthermore, one of the top five sources is listed in the AISeL, \textit{Proceedings of the International Conference on Information Systems} with 3 articles. It is noteworthy that the IEEE Xplore database does not yield a substantial number of relevant articles, either in the list of the top five most relevant sources for the relevant articles or in the set of final relevant articles. This indicates that the topic of AI reliance is currently most dominantly centered around HCI research, with only a slight touch upon core computer science research.

\section{Classification Framework for the Literature}
\label{sec:classification}

A structured framework is required to effectively review the existing literature on AI reliance. Currently, no such framework exists. However, STS theory provides a natural basis on which to develop one. This theory consists of four key components that can be used to gain insights into AI reliance research. We therefore use STS theory as the basis for deriving a framework to categorize the literature on AI reliance. Building on the components of an STS, we identified additional subcategories commonly found in AI reliance studies to help us understand the factors that influence AI reliance. In this section, we present the development of the framework. In the subsequent section, we apply this framework to classify existing AI reliance research.

\subsection{Methodology}

\subsubsection*{Framework development} The four components of an STS provide a profound perspective to consider AI reliance. These components serve as the basis for classifying the relevant articles based on their primary focus. We employed the following methodology to identify relevant subcategories. Initially, three authors screened all queried articles independently. Next, six articles were randomly selected for an in-depth read by all three authors. The focus was on the common themes, concepts, and interesting findings of these papers. Based on the initial screening and in-depth reading of the subset of articles, an author workshop was held to discuss the themes, concepts, and interesting findings. Finally, nine subcategories were identified. These nine subcategories were subsequently mapped to the four components of the STS. This approach enables the identification of common themes in AI reliance literature from the ground up, based on empirical data. Furthermore, it quickly became clear from the sample of queried articles that the common themes converge towards the identified concepts. This makes it a sound foundation on which to classify literature. However, while the subcategories converged, the characteristics diverged. Many articles in the sample had different characteristics within the subcategory. This suggests that the findings should be extended beyond the sample to include all queried articles. The classification framework is presented below. 

The goal of the classification framework is to identify an abstraction level that applies to all AI reliance literature while remaining precise enough to generate insights. It quickly becomes apparent that including concrete characteristics of specific parts of the STS, such as the technical architecture of AI systems or the age and gender of human users, would not achieve this goal. Such details are researched and manipulated in only a small subset of articles. With this in mind, the framework presented below has been developed. It can be applied to all AI reliance literature to generate insights about the examined STS.

\subsubsection*{Application of Framework} This framework is applied to all queried articles to classify, categorize, and describe them. To apply this framework, we employed qualitative coding based on the nine subcategories serving as codes to apply the framework, i.e., deductive coding \cite{saldana2021coding}. For this purpose, the qualitative coding tool MAXQDA was used. One author coded all queried articles. All other authors quality-checked a random sample of articles iteratively. Disagreements with the coding were discussed, and corrections were made retrospectively for all articles that had already been coded and integrated into the coding for all remaining articles. This ensured high-quality coding of the articles using the classification framework as a basis, as well as reliable results. Finally, the labels were manually transferred to a concept matrix spanning all articles and nine subconcepts. This concept matrix was used to present the findings in \Cref{sec:results}, albeit divided into the four components of an STS for clearer presentation.

\subsection{Classification Framework}

\begin{table}[!htbp]
    \centering
    \begin{tabularx}{\textwidth}{|c|cl|X|} \hline
        \textbf{Concept} & \multicolumn{2}{c|}{\textbf{Subconcept}} & \textbf{Description} \\ \hline
        \multirow{8}{*}{\rotatebox[origin=c]{90}{\textbf{(a) Environment}}} & (1) & \textit{Task} & Humans are confronted with a variety of decision problems. One can distinguish between decision problems with objectively correct advice, such as detecting diseases in X-ray images, and decision problems with no objectively correct advice and only subjective results, such as music preferences. \\\cline{2-4}
          & (2) & \makecell[l]{\textit{Setting}} & The manner in which data is collected for articles may vary, and this has implications for the results. Literature can be classified according to the manner in which data is collected, such as from crowdworkers, like MTurk, or from domain experts, such as medical experts. \\\cline{2-4}
          & (3) & \textit{Use cases} & Humans can receive advice in a multitude of contexts. Consequently, the present literature can be classified according to the specific use cases where an AI system is used to present advice to a human, such as in the financial or medical domain. \\ \hline
        \multirow{2}{*}{\rotatebox[origin=c]{90}{\textbf{(b) Interaction}}}  & (4) & \makecell[l]{\textit{Decision-making} \\ \textit{approach}} & There are multiple ways in which AI advice can be presented to a human decision-maker for incorporation into their decision-making approach. A distinction can be made between a single-stage decision-making approach, where AI advice is directly communicated to the user, and a two-stage decision-making approach, where the user must make an initial decision without the advice. \\ \cline{2-4}
          & (5) & \makecell[l]{\textit{Reliance} \\ \textit{measure}} & To quantify the interaction between the social and technical components, it is necessary to measure the extent to which humans rely on AI advice. This reliance can be measured using a variety of metrics that have been proposed in the literature. \\\hline
        \multirow{6}{*}{\rotatebox[origin=c]{90}{\textbf{(c) Social}}} &   (6) & \makecell[l]{\textit{User} \\ \textit{training}} & AI systems are often complex in nature, and users may not immediately understand the implications and derivation of advice. Initial training could help users understand how the AI system works and influence reliance. However, not all articles employ training, therefore, this can be used as a way to categorize the literature and distinguish between articles with and without training. \\\cline{2-4}
          & (7) & \makecell[l]{\textit{Performance} \\ \textit{feedback}} & It is common for humans to interact with AI systems on multiple occasions. This is also the case in the majority of current research, where reliance is considered over multiple task instances. Consequently, one can distinguish between articles that do not provide feedback during these instances and those that provide some feedback after a decision task. This could help users calibrate their reliance on AI advice. \\\hline
        \multirow{6}{*}{\rotatebox[origin=c]{90}{\textbf{(d) Technical}}}  & (8) & \makecell[l]{\textit{AI system} \\ \textit{implementation}} & The implementation of AI systems is typically accomplished through various algorithms and architectures. In the context of user studies, users may not necessarily interact with a live AI system. The existing literature on this topic can be broadly classified into three categories: WOZ approaches, articles that manually sample the decision instances communicated to the user, and articles where users are confronted with live AI systems. \\\cline{2-4}
          & (9) & \makecell[l]{\textit{Transparency} \\ \textit{mechanism}}& A common design choice for AI systems is the incorporation of transparency mechanisms. Consequently, articles can be classified into three categories: those that present explanations to the user on how the AI system arrived at the advice, those that present a form of performance/uncertainty of the AI system, and those that do not present any transparency mechanism to the user. \\ \hline
    \end{tabularx}
    \caption{Concepts used to classify the literature on AI reliance, based on the four components of an STS: (a) environment, (b) mutual interaction, (c) social component, and (d) technical component.}
    \label{tab:concepts}
\end{table}

For the component of the environment, we identified three subconcepts: (1) \textit{task} states whether the decision problem has objectively correct advice, i.e., ground truth, or whether the AI presents advice for a decision task with a subjective outcome, such as ethical dilemmas or personal preferences. The (2) \textit{setting} in which the data is collected may influence the reliance of human decision-makers. For example, whether the data is collected in a real-world setting or in a laboratory experiment might have an impact on user behavior. Finally, the articles can be differentiated according to their (3) \textit{use cases}. This provides a general overview of the coverage of AI reliance research in real-world problems. This may also allow for the identification of domains where AI reliance is considered more often than others.  

In addition to the environment, the STS also consists of the mutual interaction between the social and technical components. This is influenced by the (4) \textit{decision-making approach} used in the different articles. There are two main approaches: the one-step decision-making approach, where the AI advice is given to the user together with the decision task, and the two-step decision-making approach, where the human has to make an initial assessment before revising it after being confronted with the AI advice. These approaches may also influence the (5) \textit{reliance measurement}. Only unified measurements ensure comparable results and generalizable conclusions. However, AI reliance has been assessed using a variety of inconsistent measures. We therefore examine how the articles measure reliance. 

The social component relates to the user. The user can calibrate the reliance based on the experience they have with the tasks and the AI system. On the one hand, a (6) \textit{user training} can be conducted before the decision-making tasks, for instance, through trials and tutorials. This helps the user get familiar with the AI advice. On the other hand, training can also be continuous (and live) during the tasks, which involves providing (7) \textit{performance feedback} about the decisions based on the AI advice. Both approaches are considered when analyzing the literature. Both have the potential to influence the reliance of users on the system and, therefore, the social component.

The technical component refers to the AI system itself. The (8) \textit{AI system implementation} influences the AI reliance. We abstract this performance to the three cases of ``Wizard of Oz'' (WOZ) studies, manually selected labels by the authors, and a live AI system. In WOZ studies, no genuine AI system is constructed; rather, users interact with human-generated advice, which is labelled as AI advice. Further, one of the most common design decisions for AI systems, especially when user interaction is a primary concern, is the incorporation of (9) \textit{transparency mechanisms}, such as explanations. By examining these subconcepts, we can gain an understanding of the technical component. 

The framework is based on the components of an STS. The subconcepts were derived by gaining insight into the articles and extracting the most important and interesting aspects. These concepts and categorizations provide a framework for literature on AI reliance. In the following section, we present the results of categorizing the literature on these concepts and subconcepts, providing a structured overview of the current literature. 

\section{Results}
\label{sec:results}

Based on the framework introduced in the previous section (\Cref{sec:classification}), we analyze the current literature on AI reliance in this section. In \Cref{sec:settings}, we present the \textit{environment} where AI reliance occurs in the identified articles. In \Cref{sec:measure}, we present the general conceptualizations of the \textit{mutual interaction} of the technical and social components, namely, reliance, by reviewing the measures and decision-making approaches. In \Cref{sec:social}, we address the \textit{social component} of the STS and review the way users come to know the systems. Finally, to include the \textit{technical component}, we review the properties of the AI system in \Cref{sec:ai_properties}. In this section, we outline the predominant trends in the literature, highlighting potential limitations and issues that will be further explored in the following section.

\subsection{Exploring the Environment of AI Reliance}
\label{sec:settings}

The first component of the categorization framework is the environment where AI reliance is observed. This is defined by the subconcepts of the \textit{use cases} for which AI reliance is studied, the \textit{setting} in which it takes place, and the \textit{tasks} involved. The following attends to those three dimensions of the environment. 

As for the setting, most articles employ crowdworker marketplaces, such as Amazon Mechanical Turk (MTurk) or Prolific, instead of professionals or experts. Indeed, 47 articles use online crowdworker marketplaces to recruit participants. The remaining articles employ a variety of recruitment strategies, including domain experts (n = 9), students (n = 5), and convenience sampling (n = 10), without further restrictions on the participants. Overall, most articles use crowdworker marketplaces to recruit participants, and almost all of the articles rely on controlled and isolated settings. The prevalence of such isolated settings and the use of online participants pose challenges to the \textit{external validity} of these studies, which is at odds with the complex, sociotechnical nature of reliance behaviors. This also contradicts the richness of use cases depicted in the literature, raising the question of whether crowdworkers, students, and sampled non-experts can indeed provide reliable assessments in some specific use cases. 
\pgfplotstableread{
Label series1 series2 
{\footnotesize Logical decision-making} 15 1
{\footnotesize Finance \& business} 9 3
{\footnotesize Medical} 8 0
{\footnotesize General image classification} 8 0
{\footnotesize General text classification} 7 0
{\footnotesize Housing} 4 1
{\footnotesize Autonomous vehicles} 3  0 
{\footnotesize Preferences} 0 3
{\footnotesize Recidivism} 3 0
{\footnotesize Student performance} 3 0
{\footnotesize Ethical} 0  2
{\footnotesize General video classification} 2 0
{\footnotesize Labeling} 2 0
}\testdata

\begin{figure}[htbp]
    \centering
     \begin{tikzpicture}

    \begin{axis}[
        ybar stacked,
        minor y tick num = 4,
        ymin=0,
        ymax=18,
        xtick=data,
        legend style={cells={anchor=west}, legend pos=north east},
        reverse legend=false, 
        xticklabels from table={\testdata}{Label},
        xticklabel style={text width=1.75cm,align=right,xshift=0pt, rotate=90},
        height = 4.5 cm, 
        width = 16 cm,
        grid,
    ]
    \addplot [fill={rgb,255:red,169;green,169;blue,255}] table [y=series1, meta=Label, x expr=\coordindex] {\testdata};
    \addlegendentry{objective}
    
    \addplot [fill={rgb,255:red,238;green,216;blue,190}] table [y=series2, meta=Label, x expr=\coordindex] {\testdata};
    \addlegendentry{subjective}
    \addplot [
        ybar, 
        nodes near coords,
        nodes near coords style={%
            anchor=south,%
        },
    ] table [ y expr=0.00001, x expr=\coordindex] {\testdata};

    \end{axis}
    \end{tikzpicture}
    \caption{Overview of all use case categories and categorization of whether the goal of the task has no objective correct advice or whether advice can only be subjective. Some articles present tasks from multiple categories and are therefore listed multiple times.}
    \Description{The figure shows a bar chart for the use case categories and the categorization of whether the goal of the task has no objective correct advice or whether advice can only be subjective. The numbers are as follows. The Logical decision-making group has 15 objectives and one subjective goal. The Finance \& business group has nine objectives and three subjective goals. The medical has eight objective and zero subjective goals. The General image classification has eight objective and zero subjective goals. The General text classification has seven objective and zero subjective goals. The Housing group has four objectives and one subjective task. The Autonomous vehicles group has three objectives and zero subjective goals. The Preferences group has zero objective and three subjective goals. The Recidivism group has three objective and zero subjective goals. The Student performance group has three objective and zero subjective goals. The Ethical group has zero objective and two subjective goals. The General video classification has two objective and zero subjective goals. The Labeling has two objective and zero subjective goals.}
    \label{fig:use_cases}
\end{figure}

\Cref{fig:use_cases} presents an overview of the \textit{use cases} covered in the studied papers. For a more in-depth analysis, we propose to divide the use cases into two groups: those where the outcome is of a subjective nature, linked to user preference, and those where the AI aims to present objectively correct advice, meaning use cases with a ground truth present. Both require fundamentally different AI systems, with different goals. While the first-category systems aim to estimate the individual user preferences, the systems focusing on objective outcomes can be trained with ground truth data. As both groups of use cases possess different characteristics, it is important to distinguish these cases and become aware of the specific use case that the AI system aims to address. 

Overall, there is a range of studies in which the system is said to estimate the user's subjective preference. A prime example of the group of subjective use cases is \textit{recommender systems}, such as music recommendations \cite{logg_algorithm_2019, radensky_i_2023} or recommendations on the attractiveness of other people \cite{rhue_beautys_2019}. In these cases, the AI system cannot be trained on nor predict an objective ground truth (because there is none) but must rather aim at estimating the user's preferences. Further subjective outcomes are also found in use cases where one might not expect them at the outset. For example, managers may have preferences about investment opportunities \cite{keding_managerial_2021} or customers may have preferences about life insurance products \cite{bertrand_questioning_2023}. Some research even frames stock trading use cases as a task with subjective outcomes, as some stocks may be preferred over others \cite{castelo_task-dependent_2019}. There are several use cases with less clear subjective characteristics that are nonetheless declared subjective in the analyzed articles. These use cases include estimating the percentage of the general public who will find a specific painting creative \cite{hou_who_2021}. While there might be no definitive ``right answer'' for a specific painting, estimating public preference does not really address a user's subjective opinion, but rather their ability to make a good guess. Similarly, ethical decision-making is often presented as subjective, such as ethical dilemmas related to military defense and rescue actions  \cite{tolmeijer_capable_2022} or a decision about a recipient for a donor kidney, where multiple viable options exist \cite{narayanan_how_2023}. Although presented as \textit{subjective}, these studies revolve around questions of what is right or wrong, where users may have different references beyond their own personal subjective assessments or wants. 

Overall, we observe that there are varieties of what might be considered subjective, starting with use cases focusing on personal preferences concerning music or attractiveness, up to decisions involving trade-offs such as investments or tenant selection, up to decisions using reference frameworks such as the general public's preferences or ethical dilemmas. We refer to this tendency in the literature as \textit{subjectivity dilution}. 

Still, the majority of the analyzed literature is concerned with use cases that have an objective ground truth. Some of these cases are anchored in real-life scenarios, while others are isolated experiments. It is also noteworthy that in certain use cases, the objective is to predict future events, while other use cases are concerned with directly uncovering the ground truth without the need for such delayed labels that only uncover at a later point in time. 

In the use case category of finance \& business settings, numerous articles are concerned with loan applications \cite{dikmen_effects_2022, he_how_2023, purificato_use_2023, jakubik2022empirical} and stock trading \cite{chacon_longitudinal_2022, chua_ai-enabled_2023, cau_supporting_2023}. For instance, non-expert users are asked to decide whether to grant a loan in a peer-to-peer setting based on provided information about the loan applicant. Their decisions are then compared with delayed labels from ground-truth data regarding the applicant's actual loan performance \cite{dikmen_effects_2022}. Loan applications and stock trading appear to be favored use cases in the current literature. One reason may be the close proximity to other machine learning research, which also frequently considers these cases and provides labelled data. In addition to loan applications and stock trading, other related use cases include basketball betting \cite{elder_knowing_2022} and the prediction of incoming call center calls \cite{berger_watch_2021}. 

In contrast to future-oriented predictions, some settings rely on real-time assessment. This is the case in the use case category of housing, where users should estimate the prices of houses \cite{chiang_youd_2021, chiang_exploring_2022, prabhudesai_understanding_2023} or find a fitting flat using online platforms \cite{gupta_trust_2022}. Another domain frequently explored in the context of AI reliance is the medical field. The specific applications fields can be grouped into patient and elderly care \cite{zukerman_influence_2023, lee_understanding_2023}, image classification, such as MRI and X-ray \cite{cabitza_ai_2023, fogliato_who_2022}, and disease detection, where various diseases are considered, such as diabetes risk \cite{du_role_2022}, myocardial infarction, heart conditions \cite{panigutti_understanding_2022}, cancer risk assessment \cite{wysocki_assessing_2023}, or sepsis classification \cite{barr_kumarakulasinghe_evaluating_2020}. In general, these articles also opted for experts, such as doctors, as decision-makers, as opposed to crowdworkers. This can be attributed to the high specialization required to detect diseases. The last groups of real-life anchored use cases are \textit{autonomous vehicles} \cite{srivastava_improving_2022, ajenaghughrure_psychophysiological_2021, okamura_adaptive_2020}, \textit{student performance} assessment \cite{zhao_evaluating_2023, rastogi_deciding_2022,dietvorst_overcoming_2018}, and the case of \textit{recidivism} assessment, often based on the COMPAS algorithm \cite{chiang_are_2023, grgic-hlaca_human_2019, wang_are_2021}. In summary, we identified various settings that explore tasks closely resembling real-life scenarios, with finance, business, and medicine being the most dominant. While future-oriented predictions differ significantly from assessments in real life, they all involve uncertainty as a core element. Yet, in the analyzed literature, these are often presented as tasks with a clear, precise answer, thus diverging from real-life settings in a critical aspect.

In contrast to the decision-making task described above, which involves a high level of uncertainty, isolated experiments require users to make logical decisions or solve simple classification problems. Logical decision-making is the most diverse category and includes counting tasks \cite{bogert_humans_2021},  and games, such as chess \cite{bayer_role_2022}, maze-solving \cite{vasconcelos_explanations_2023}, or path-building \cite{vered_effects_2023}, where the AI presents advice on the correct strategy. Finally, some articles address general image, text, or video classification tasks. Image classification tasks include noisy images based on the ImageNet dataset \cite{tejeda_ai-assisted_2022, lemus2022empirical}, handwritten images from the MNIST dataset \cite{cau_effects_2023}, bird images \cite{lu_strategic_2023, schemmer2023towards, cabrera_improving_2023}, or satellite images \cite{cabrera_improving_2023, morrison_evaluating_2023}. The text classification tasks include sentiment analysis of reviews \cite{cau_effects_2023, bansal_does_2021, lai_selective_2023} or fake review detection \cite{schemmer_appropriate_2023,cabrera_improving_2023} and review credibility assessment \cite{robbemond_understanding_2022}, question answering tasks \cite{bansal_does_2021,goyal-etal-2023-else}, or translation tasks \citet{mehandru-etal-2023-physician}. Finally, for video classification, video clips should be categorized \cite{kim_learn_2021} or activities should be recognized \cite{nourani_anchoring_2021}. In summary, additional to the real-world, anchored use cases, we observe that many articles are concerned with isolated experiments. Rather than providing insight into the specific domains in which AI systems are relied upon, these cases are employed to gain a general understanding of AI reliance. 

Other potential characteristics of a task---such as task interdependency, priority, purpose, scope, resource requirements, or completion conditions---and their influence on AI reliance remain largely unexplored in the queried articles. Only the complexity of the task has been explicitly studied. The research demonstrates that users tend to rely on AI more for complex tasks than for simpler ones \cite{bogert_humans_2021}, and that the effectiveness of uncertainty visualizations in supporting appropriate reliance depends on task complexity \cite{zhao_evaluating_2023}. This indicates that task characteristics can significantly impact reliance behaviors and call for more research in this direction. We refer to this shortcoming of the existing research as \textit{task characteristics neglect}. 

In summary, analysis of various use cases, settings, and specific tasks reveals that a significant portion of the queried articles use experiments and data collection methods that may raise concerns about their external validity. Besides clearly constructed experiments designed not to resemble real-world settings, such as those involving logical decision-making or simple classification tasks, even studies that attempt to address external validity by introducing more complex, real-life decision-making situations have limitations. They misrepresent tasks reliant on best-guess approaches or social reference frames as subjective. They also imply the existence of a definitive truth in situations characterized by high uncertainty and frequent influence of factors not typically represented in the data or easily controlled. For example, defaulting on a loan is often caused by external factors, with machine learning algorithms and professionals frequently relying on proxies to predict the likelihood of such factors. Moreover, many of these tasks are assigned to non-experts and inexperienced users from outside the specific domain and delivered as online surveys, rather than as part of a more overarching process typically present in real life. Overall, while we acknowledge significant effort in trying to resemble real-life decisions, we observe a lack of studies that adequately address all of these aspects. We refer to this observation as \textit{external validity deficit}.

\subsection{Measuring the Interaction of Human and AI}
\label{sec:measure}

We examine the interaction between the social and the technical components of the STS by reviewing \textit{approaches} and \textit{measures} used to quantify AI reliance. We begin by presenting the approaches employed in the experiments, forming a higher-level perspective on the experimental designs. Later, we will focus on the measurements used within those experiments as a lower-level perspective.

\subsubsection*{Decision-Making Approach}
As described in \Cref{sec:related_work}, AI advice can either be presented before a decision (single-stage decision-making approach) or after an initial decision by the human (two-stage decision-making approach). Generally, in the \textit{single-stage decision-making} approach, users are directly confronted with AI advice without an initial decision. The final decision from the human-AI interaction is recorded only after the user has been presented with the AI advice. Nonetheless, in the articles, there is some variance concerning the specific approaches. For instance, one article measures the human decision twice after being exposed to the AI decision in the context of music recommendations \cite{radensky_i_2023} to account for potential familiarity effects. Further variances include the binary decision of accepting or rejecting the AI advice \cite[e.g.][]{bertrand_questioning_2023}, or the decisions for delegating certain tasks\cite[e.g.][]{okamura_adaptive_2020}. Another variant is that the AI advice is implemented automatically until a human intervention, as in the case of automatic driving games \cite[e.g.][]{ajenaghughrure_psychophysiological_2021}. Despite this variety in single-stage decision-making approaches, we found no study comparing these specific designs. For example, comparing different decision types, such as choosing an answer versus choosing whether to accept an AI's suggestion, could reveal more about the reliance behavior. In summary, while single-stage decision-making approaches share similarities, careful consideration reveals potential improvements that could allow for the aggregation and comparability of different study designs.

Many articles employ a two-stage decision-making approach. They explicitly collect information about the initial human decision. Yet, also within this category, we find some varieties worth reporting. A notable example is an article focusing on the ethical dilemma of selecting a kidney transplant recipient based on three factors: wait time, prior donor status, and disease stage \cite{narayanan_how_2023}. Based on a set of initial assessments shown without AI recommendation, the system determines a user's preferences regarding the importance of those factors and presents this information to the user, along with the preferences of an AI that will be used for making recommendations. Subsequently, the user is asked to select a kidney recipient again, this time with the AI's recommendation visible. Using this design, the authors aimed to understand how information about the ethical preferences implemented in the AI, in relation to one's own ethical preferences, impacts AI reliance. Some studies also use more than two stages to study reliance. \citet{morrison_evaluating_2023}) employ a multistage approach that presents more and more explanations to the users, and the user can adapt their decision after each step. Conversely, \citet{elder_knowing_2022} does not explicitly collect an initial human decision, but introduces a waiting period before exposing the human to the AI advice, which also aims to enforce this two-stage decision-making approach.  In one example, the two-stage decision-making approach is only made during training \cite{chiang_youd_2021}. With the exception of these few instances, all other articles employ a two-stage decision-making approach, where they explicitly collect the initial human decision and allow the human to revise after exposure to the AI advice. 

Overall, we find a nearly even split between articles that have this two-stage decision-making approach (n = 36) and those with a single-stage decision-making approach, directly presenting AI advice without collecting the initial human decision (n = 35)\footnote{One article explicitly tests both \cite{fogliato_who_2022}, which we count toward the set of articles with a two-stage decision-making approach}. The even split in the literature indicates that there is a potential tension between experimental precision, which aims at identifying and quantifying the effective impact of AI, and the realism of the decision approach. In the one-stage approach, the impact of AI is hard to identify, as it remains unclear if the user acts on their internal beliefs or on the AI's advice, and whether they considered AI's advice at all. On the other hand, the two-stage decision-making approach adds unnatural complexity to the decision-making process, which is very unlikely in real-world applications, which frequently deliver AI advice and leave it open to the user to use or ignore it. We refer to this observation as \textit{precision-realism tension}. The additional variety of approaches within the two classes further complicates the situation, as does the variety of metrics and measurements described below.

\subsubsection*{Metrics and Measures}

\begin{figure}
    \pgfplotstableread{
    Label series1 series2 
    {\footnotesize Agreement percentage} 16 13
    {\footnotesize \textit{Weight of advice}} 0 12
    {\footnotesize \textit{Switch percentage}} 0 11
    {\footnotesize Survey items} 7 3
    {\footnotesize Accuracy} 6 1
    {\footnotesize Manual override} 2 1 
    {\footnotesize \textit{Appropriateness of reliance}} 0 2 
    {\footnotesize Qualitative statements} 2 0
    {\footnotesize Delegation} 1 1
    {\footnotesize Others} 2 3
    }\testdata
    \centering
    \begin{tikzpicture}
        \begin{axis}[
        ybar,
        minor y tick num = 4,
        ymin=0,
        ymax=18,
        xtick=data,
        legend style={cells={anchor=west}, legend pos=north east},
        reverse legend=false, 
        xticklabels from table={\testdata}{Label},
        xticklabel style={text width=1.4cm,align=right,xshift=0pt, rotate=90},
        height = 5 cm, 
        width = 16 cm,
        single ybar legend,
        grid]
        \addplot [fill={rgb,255:red,169;green,169;blue,255}, nodes near coords] table [y=series1, meta=Label, x expr=\coordindex] {\testdata};
        \addlegendentry{single-stage}
        \addplot [fill={rgb,255:red,238;green,216;blue,190}, nodes near coords] table [y=series2, meta=Label, x expr=\coordindex] {\testdata};
        \addlegendentry{two-stage}
        \end{axis}
    \end{tikzpicture}
    \caption{Reliance measures, with some articles containing multiple measures. The \textit{italics} measures strictly require a two-stage decision-making approach.}
    \Description{This figure shows a bar chart for the different measures and the underlying decision-making approach. The numbers are as follows. The Agreement percentage is applied by 16 articles in a single-stage decision-making approach and by 13 articles in a two-stage decision-making approach. The Weight of advice is applied by zero articles in a single-stage decision-making approach, and by twelve articles in a two-stage decision-making approach. The Switch percentage is applied by zero articles in a single-stage decision-making approach and by eleven articles in a two-stage decision-making approach. The Survey items are applied by seven articles in a single-stage decision-making approach and by three articles in a two-stage decision-making approach. The Accuracy is applied by six articles in a single-stage decision-making approach and by one article in a two-stage decision-making approach. The Manual override is applied by two articles in a single-stage decision-making approach and by one article in a two-stage decision-making approach. The Appropriateness of reliance is applied by zero articles in a single-stage decision-making approach and by two articles in a two-stage decision-making approach. The Qualitative statements are applied by two articles in a single-stage decision-making approach and by zero articles in a two-stage decision-making approach. The Delegation is applied by one article in a single-stage decision-making approach and by one article in a two-stage decision-making approach. Other measures are applied by two articles in a single-stage decision-making approach and by three articles in a two-stage decision-making approach.}
    \label{fig:measures}
\end{figure}

Existing literature makes it evident that a variety of metrics are used to assess the degree of reliance. Some of these metrics align with the decision-making approaches introduced above.
In the single-stage decision-making approach, the number of agreements between human decisions and AI advice can be counted. In the two-stage decision-making approach, the number of switches from the initial human decision to the AI advice can also be counted. Nevertheless, numerous additional metrics are employed in the literature, which are summarized in \Cref{fig:measures}. Furthermore, the formulas for quantitative measures are presented in the digital appendix.

The most common measure of AI reliance is also one of the simplest, the \textit{agreement percentage}. It expresses how many decision instances a user has the same answer as the AI. It can be measured in both settings, the single-stage \cite[e.g.][]{tolmeijer_capable_2022, jakubik2022empirical, zhao_evaluating_2023} and the two-stage \cite[e.g.][]{chen_understanding_2023, he_how_2023, lu_human_2021}. It is, therefore, a convenient and intuitive measure, but some researchers argue that this match is not enough to measure reliance \cite{schmitt_towards_2021}. A downside might be that instances where the human simply agrees with the AI advice, but would have decided the same even without the AI advice, are also counted toward this measure. Some articles, therefore, require a harder condition for reliance.

The \textit{switch percentage} aims at this condition, as it measures the percentage of instances where the decision-maker switched their decision to the AI advice after being confronted with it. Interestingly, most articles that use switch percentage as a measure also use agreement percentage \cite[e.g.][]{lu_strategic_2023, srivastava_improving_2022, he_how_2023}. Only a few articles exclusively use switch percentages \cite{narayanan_how_2023, schmitt_towards_2021, grgic-hlaca_human_2019}. This shows that the agreement percentage is also seen as a viable measure. This approach has another advantage. If the use case is not a discrete decision case, but rather a regression case, the \textit{WOA} (\Cref{sec:related_work}) can be used. Note that if the estimates and advice are binary, the WOA and switch percentage align. In other cases, WOA can give an indication of the strength of the reliance, referring to how much the AI advice influenced the decision-maker's final decision. The idea of switching the decision was also used to define the \textit{appropriateness of reliance} \cite{schemmer_appropriate_2023, schemmer2023towards} that counts the cases where reliance was appropriate and ignores overreliance and underreliance. 

Another way reliance is sometimes measured is by using task \textit{accuracy} as a proxy \cite[e.g.][]{cabrera_improving_2023, rastogi_deciding_2022, bansal_does_2021}. In a within-subject design, participants make decisions with and without AI advice. If the performance of the final decisions improves, the participant is said to rely on AI advice. This is often linked to the concept of complementary team performance, where a human-AI team outperforms the individual members alone, making accuracy a measure of appropriate reliance. Some articles define AI reliance as \textit{non-action}, with users intervening only when prompted by the AI system, and non-prompted intervention is defined as non-reliance \cite{ajenaghughrure_psychophysiological_2021, zukerman_influence_2023, elder_knowing_2022}.  Similarly, sometimes delegation is also measured as reliance \cite{okamura_adaptive_2020, chiang_youd_2021}. 

Some articles employ tailored measures. For instance, \citet{kim_learn_2021} assesses reliance based on the frequency with which a user requests XAI assistance. Other examples use eye gaze as a measure of AI reliance \citet{cao_how_2023}, identify patterns in user behavior to quantify reliance \cite{cabitza_ai_2023}, or use regression analysis \cite{fogliato_who_2022}. While these measures offer insights into AI reliance and attempt to reduce dependency on explicit decisions and user statements---potentially mitigating manipulation effects---they are context-specific and may not be applicable to other studies. Furthermore, some of the measurements make strong assumptions to argue for their applicability, where unintentional behaviors like mouse movements are said to link with agreement or disagreement.

While the above measures depend on observing user behavior, other articles use subjective measures or combine them with objective ones. Several articles employ survey items, typically on a 5- or 7-point Likert scale, asking participants the extent to which they relied on the AI's advice. Most of these articles exclusively use a survey to measure reliance \cite[e.g.][]{purificato_use_2023, chua_ai-enabled_2023, nourani_anchoring_2021}, while others combine surveys with metrics like gaze duration or perception-based measures \cite{lee_understanding_2023, cao_understanding_2022}. Many articles use newly developed items such as ``I relied on assessment scores and analysis from the system for my final assessment'' \cite{lee_understanding_2023} or ``I relied on the AI suggestion in the previous task'' \cite{cao_understanding_2022}. Others employ proxy metrics, such as \textit{intention to accept AI-based recommendations} \cite{chua_ai-enabled_2023}, or adapt existing metrics \cite{purificato_use_2023}. Overall, the analysis reveals limitations in self-report-based instruments, prompting researchers to use alternative strategies and highlighting the need for a more systematic approach to develop a precise scale. Two articles use a ranking of AI system interface features, with the highest-ranked feature being said to be relied upon by users \cite{cau_effects_2023, cau_supporting_2023}, making this a self-report rather than a behavioral measure.  Other self-report measures are \textit{qualitative statements} from interviews  \cite{bertrand_questioning_2023, prabhudesai_understanding_2023}, aiming to understand the internal emotional or cognitive processes that lead to reliance.

The previously introduced nature of the task does not directly relate to the use of subjective and objective measures. For example, reliance on subjective tasks---such as accepting a music track recommendation \cite{radensky_i_2023}---can be measured using behavioral, objective metrics like the number of times a user follows the recommendation. Conversely, tasks utilizing objective ground truth, like loan application assessments, may employ self-reports and other subjective approaches to assess reliance \cite{purificato_use_2023}. These aspects are independent and can be freely combined.

Overall, there are evident clusters of measures that are employed with greater frequency. The most prevalent measure is the agreement and switch percentage. On occasion, surveys or qualitative statements are used as an additional indicator of reliance. This indicates that, first, there is no clear consensus on how to measure reliance, and second, reliance is a complex construct, and the use of a single measure is not always sufficient. This further contributes to the precision-related challenges described above, as well as to the conflict over the definition of reliance as a behavior: is it the adaptation of a decision based on an AI's recommendation, or is it any behavior related to following or ignoring an AI's advice? We refer to this dilemma as \textit{measures discrepancy}. We will examine the implications of various metrics being employed and the tensions between them in greater detail \Cref{sec:critical_review}. 

\subsection{Human User as the Social Component}
\label{sec:social}

After establishing settings,  use cases, and measures of reliance, we proceed to the social component. While many user attributes could influence their reliance behavior, including demographics, psychological profile, or expertise, studies focusing on such relationships are rare. This creates a significant void that deserves further attention. Out of the analyzed studies, none attends to the relationship between typical demographic characteristics like age or gender on reliance. As for individual characteristics' impact on reliance behavior, we identified individual studies concerning the impact of expertise \cite{schemmer2023towards, bayer_role_2022, dikmen_effects_2022} and confidence \cite{cau_supporting_2023} on AI reliance. The studies indicate, for example, that explanations are useful for expert users in achieving appropriate reliance but are irrelevant for non-experts \cite{bayer_role_2022} or that a user's general confidence in AI moderates which forms of explanations improve task performance \cite{cau_supporting_2023}. This suggests that various variables can have both direct and moderating effects, indicating complex interactions between them. However, the analyzed studies overlook other individual characteristics, a phenomenon we refer to as \textit{user characteristics neglect}.

Nevertheless, one aspect emerges as interesting from the analysis: the impact of a user's experience with the specific AI system. In many decision-making tasks, humans are confronted with multiple decision instances in rapid succession. For instance, a doctor diagnoses numerous patients. Experience gained from one instance may influence decisions in the next.
\Cref{tab:reliance_change} distinguishes between training with the AI system and performance feedback during the main task. Not all articles explicitly state whether they provide feedback and/or training. Consequently, when feedback and/or training information was absent, we classified it as \textit{no feedback and/or training}. In some tasks, this information might be implicit, such as an autonomous vehicle driving game \cite{ajenaghughrure_psychophysiological_2021}, where no crash with the autonomous vehicle indicates that the AI provided positive advice, and was thus inferred accordingly. 
 
\begin{table}[htbp]
    \centering
    \begin{tabularx}{\textwidth}{rcc} \toprule
         & \textbf{Feedback on AI performance} & \textbf{No feedback on AI performance} \\ \midrule
       \textbf{Training with the AI}  & 7 & 16  \\
       \textbf{No training with AI}  & 10  & 38 \\ \bottomrule
    \end{tabularx}
    \caption{Two-dimensional matrix on AI training and feedback.}
    \label{tab:reliance_change}
\end{table}

As illustrated in \Cref{tab:reliance_change}, most articles do not provide training or feedback to users, possibly due to under-reporting in the article or the ease of data collection through such an experimental design. Additionally, most articles do not examine changes in reliance over time or across multiple interactions, so that feedback would not have any impact on the empirical results. In certain instances, the provision of training and feedback may be impracticable, such as one-off sessions \cite[e.g.][]{bertrand_questioning_2023}. 

Some articles provide training before the main task without the AI advice \cite{schemmer2023towards} to assess the participants' skills before the main task, which could be a potential confounder of reliance. In general, most articles employ a straightforward experimental design with no training and/or feedback. This approach is employed to measure the average reliance of the participants and to allow for comparisons between groups. 

Another group of articles provides training to participants, but no feedback. This is often done to reduce potential misunderstandings and to help users understand the function of the AI system. This training allows the user to interact with the system's advice without including those interactions in the analysis. This training is sometimes limited to a single decision instance \cite{bucinca_trust_2021}, but often involves many interactions with the AI system, e.g., ten \cite{chiang_youd_2021}. In some instances, the initial training also serves as an initial screening and selection of participants \cite{fogliato_who_2022}. In summary, several articles provide training with the AI system that does not count toward the main task and final analysis. 

Some articles provide feedback, but no training. The user can learn as they interact with the AI, which can be seen as continuous training on the task. However, all decision instances count toward the overall performance. While most articles provide feedback after each decision, some provide feedback halfway through the decision instances \cite{zukerman_influence_2023, dietvorst_overcoming_2018}. In some cases, articles do not explicitly state that feedback is provided, but it can be inferred implicitly, such as the case of music recommendations \cite{radensky_i_2023}, where a participant would know about the performance of the AI model based on the recommended music.

Finally, a few articles use both training and performance feedback during the main task. Although rarely applied, this experimental design may better reflect real-world scenarios. For example, a medical professional would receive an introduction/training session with the tool and receive ongoing performance feedback by observing the patients and the system's diagnosis. There, those articles that use this experiment design are mostly articles with real-world use cases, such as the task of predicting the incoming calls of a business center \cite{berger_watch_2021}, or the case of driving an autonomous car \cite{ajenaghughrure_psychophysiological_2021}. 

Most measures of AI reliance are calculated by averaging multiple interactions of a user with the system. For example, if a user agrees with the AI system in eight out of ten interactions, the agreement percentage is 80\%. However, this ignores the effects of repeated interactions with the AI system or prior training, which might influence reliance. Yet, only a few articles explicitly measure change in reliance. For example, \citet{grgic-hlaca_human_2019} studies the COMPAS algorithm and explicitly argues with real-world judges who also have the chance to adapt their reliance, and \citet{leffrang_people_2023} investigates how users recover from bad advice given by the AI system. However, these cases are rare. 

Overall, while many articles employ training or feedback to account for potential effects of users' experience with the AI system and its performance, the majority do not explicitly address users' characteristics. Additionally, potential long-term effects are entirely absent. This exacerbates the previously mentioned external validity challenges: many tasks used in the experiments simulate situations that will likely occur multiple times, but might be distributed over time rather than concentrated in a short experimental session. This \textit{temporal effects neglect} ignores natural human traits, such as the tendency to build habits, establish cognitive shortcuts, or forget.  

\subsection{AI System as the Technical Component}
\label{sec:ai_properties}
In addition to the social component, the technical component plays a key role in STSs. It enters reciprocal interactions with the social component, leading to mutual adaptation over time. This section reviews the various AI systems used in the AI reliance literature, focusing on the nature of the AI system itself and transparency mechanisms. While many system properties could impact reliance, reliance research most prominently focuses on the interaction between transparency or explainability features and reliance. Further, most research uses experimental designs that do not require a fully functional system.

With regard to the AI system itself, many articles do not use real AI systems. Instead, users interact with mockups and dummies in WOZ studies (n = 29) \cite[e.g.][]{schmitt_towards_2021, radensky_i_2023, tolmeijer_capable_2022}, where users are told that they receive AI advice, but in fact the researchers constructed it. This methodology is commonly employed, particularly in HCI research, and streamlines research, as no AI system needs to be constructed. It is, however, crucial to exercise caution, as mocked systems do not provide real AI advice and may be biased in their output. Some of these studies even provide advice that is always correct \citet{srivastava_improving_2022, panigutti_understanding_2022}, making measuring overreliance impossible.

Some articles construct real AI systems but present manually selected instances of AI advice, i.e., manually sample the dataset (n = 19) \cite[e.g.][]{chen_understanding_2023, wang_are_2021, jakubik2022empirical}. This approach is frequently employed to achieve a specific performance of the model. Compared to WOZ designs, this approach is closer to reality, as a real AI system created the advice. Yet, hand-picked examples have downsides as real randomness is absent, even in cases where outliers are explicitly studied \cite[e.g.][]{leffrang_people_2023}.

These manually selected labels also distinguish between the model accuracy of the underlying model and the sampling accuracy of the samples presented to the decision-makers. For example, \citet{chen_understanding_2023} presents eight decision instances, five of which have correct advice, resulting in a sampling accuracy of 62.5\%, but the underlying models have accuracy above that. This distinction between sampling and model accuracy might induce biases. Only when decision-makers interact with real AI systems does sampling accuracy equal model accuracy. 

However, only a handful of articles allow users to interact with real AI systems (n = 21) \cite{schemmer_appropriate_2023, cabitza_ai_2023, purificato_use_2023}, i.e., deliver actual AI recommendations as part of the decision process. In most cases, this results in a random sample from a list of AI advice.  Many studies do not use real AI systems but rather either fully mocked interfaces or hand-picked AI advice. To address the issue, \citet{ashktorab_ai-assisted_2021} tests both a real AI and a WOZ study with a ``perfect'' AI system that always outputs correct advice\footnote{One article did not provide conclusive information about the implementation of the AI system and could not be categorized \cite{castelo_task-dependent_2019}}. Whereas this allows for more control and supports internal validity, those study designs again compromise on external validity and also limit the potential to explore previously unexplored aspects, such as an error message or incomprehensible advice. Such situations might have radical effects on reliance, but insights in this regard remain unstudied in current literature.

With regard to the characteristics of AI, research primarily focuses on explainability and transparency. For this survey, we distinguish between explanations and statements of AI performance or certainty as another form of transparency. Among the analyzed articles, 46 articles study whether providing information about AI uncertainty, performance, and confidence affects reliance \cite[e.g.,][]{cau_effects_2023, cau_supporting_2023, rhue_beautys_2019, zhao_evaluating_2023, zhang_you_2022}, or how different types of explanations influence reliance \cite[e.g.,][]{wysocki_assessing_2023, vered_effects_2023, panigutti_understanding_2022}. Some studies examine specific modalities of explanation, e.g., whether they are provided as text or graphic \cite{robbemond_understanding_2022, gupta_trust_2022}, while others focus on different explanation mechanisms, including example-based, feature-based \cite{bertrand_questioning_2023}, and counterfactual explanations \cite{scharowski_exploring_2023, lee_understanding_2023}. Only a few studies, however, explore the differences between these specific mechanisms \cite{chen_understanding_2023, du_role_2022}. 

We find that many articles (n = 27) focus on providing explanations to users.  The literature contains a variety of different types of explanations. These include feature-based, example-based, and counterfactual explanations, which are presented using different modalities, as explained below. A common type of explanation is feature-based explanations \cite[e.g.,][]{bertrand_questioning_2023}, which highlight the contribution of different model features to the specific prediction and aim to explain the model output in that way. Popular methods include SHAP values \cite[e.g.,][]{dikmen_effects_2022, du_role_2022} or the LIME method. \cite{barr_kumarakulasinghe_evaluating_2020}. Another type of explanation is example-based, where specific examples relevant to the prediction are shown to the user in an attempt to explain the inner workings of the model \cite[e.g.,][]{bertrand_questioning_2023, chen_understanding_2023}. A third type of explanation is counterfactual explanations \cite{scharowski_exploring_2023, lee_understanding_2023}. These explanations demonstrate how the model works by showcasing ``what if'' scenarios, i.e., by showing what would happen if the input to the model changed. In fact, some articles, such as \cite{chen_understanding_2023, du_role_2022, bertrand_questioning_2023}, explicitly address the impact of different explanations on reliance. However, such articles are rare, so an overview of design features that potentially impact reliance is lacking. Finally, rather than exploring different types of explanations, some articles explore specific modalities of explanation. Typically, explanations are presented in text form, but their influence may differ if they are presented in graphic form instead of text form \cite{robbemond_understanding_2022, gupta_trust_2022}. Currently, modalities are limited to visual or textual explanations, presenting an opportunity to explore various other modalities, such as auditory explanations.

In addition, in some articles, the user is provided a statement on the overall AI performance or the certainty on a specific decision task (n = 11) \cite[e.g.][]{prabhudesai_understanding_2023,rhue_beautys_2019,radensky_i_2023}. Interestingly, only a few articles present both to the user (n = 8) \cite[e.g.][]{cabrera_improving_2023,rastogi_deciding_2022,kim_learn_2021}. However, none of the articles attends to the question of whether the provided information is understandable to the user and if they interpret it correctly, leaving it open 

Finally, roughly one third of the articles do not employ any transparency mechanisms (n=25) \cite[e.g.][]{gupta_trust_2022,lu_human_2021, cao_how_2023}. These articles often focus more on general applicability, such as the case of an AI that predicts preferences, where this premise is already the subject of research \cite[e.g.][]{logg_algorithm_2019, tolmeijer_capable_2022}. Other examples are articles that are interested in different aspects, such as the above-mentioned change of reliance over time \cite{grgic-hlaca_human_2019} or the use of multiple users \cite{chiang_are_2023}. It becomes apparent that a clear majority of articles present some form of transparency mechanism, highlighting their technology-centricity. Changes to the design of the technological system are often considered more interesting than the actual users, who often consist of crowdworkers and non-specific user groups.

Many articles focus on explainable AI and mechanisms related to it. Most articles indicate that explanations and transparency mechanisms are helpful in influencing AI reliance. However, a recent meta-study finds that the impact of explanations on human–AI decision-making is inconclusive in the current literature \cite{schemmer2022meta}. Therefore, we argue that a comprehensive comprehensive, sociotechnical perspective is needed in AI reliance research. Going into even more specific details on explainability mechanisms than laid down above would exceed the scope and goal of this study, but we invite readers to conduct a systematic meta-study on the impact of explainability on reliance while also considering the other components of the STS.

Summarizing the current research on the technical component, we find that several articles do not present users with actual AI systems, but instead use the WOZ methodology. Only a few articles involve users interacting with genuine AI systems. Additionally, we observe that many articles implement transparency mechanisms for their AI systems, which may influence user reliance. Still, further mechanisms and attributes of the systems remain absent in most studies – we refer to this as \textit{explainability fixation}. Also, study designs that employ a multi-factorial design are missing, which makes it difficult to assess, e.g., how explanations and users' experience interact with each other in the context of reliance. We refer to this observation as \textit{multi-factorial effects neglect}. The implications are discussed in \Cref{sec:critical_review}.

\section{Discussion}
\label{sec:critical_review}

After a comprehensive review of the existing literature on AI reliance, we will first propose a morphological box for AI Reliance in research and practice based on what this study has uncovered as established knowledge. A second section then proposes a research agenda based on uncovered shortcomings and opportunities. 

\subsection{A Morphological Box for Sound AI Reliance Research}

\begin{figure}
    \centering
    \includegraphics[width = 0.8\textwidth]{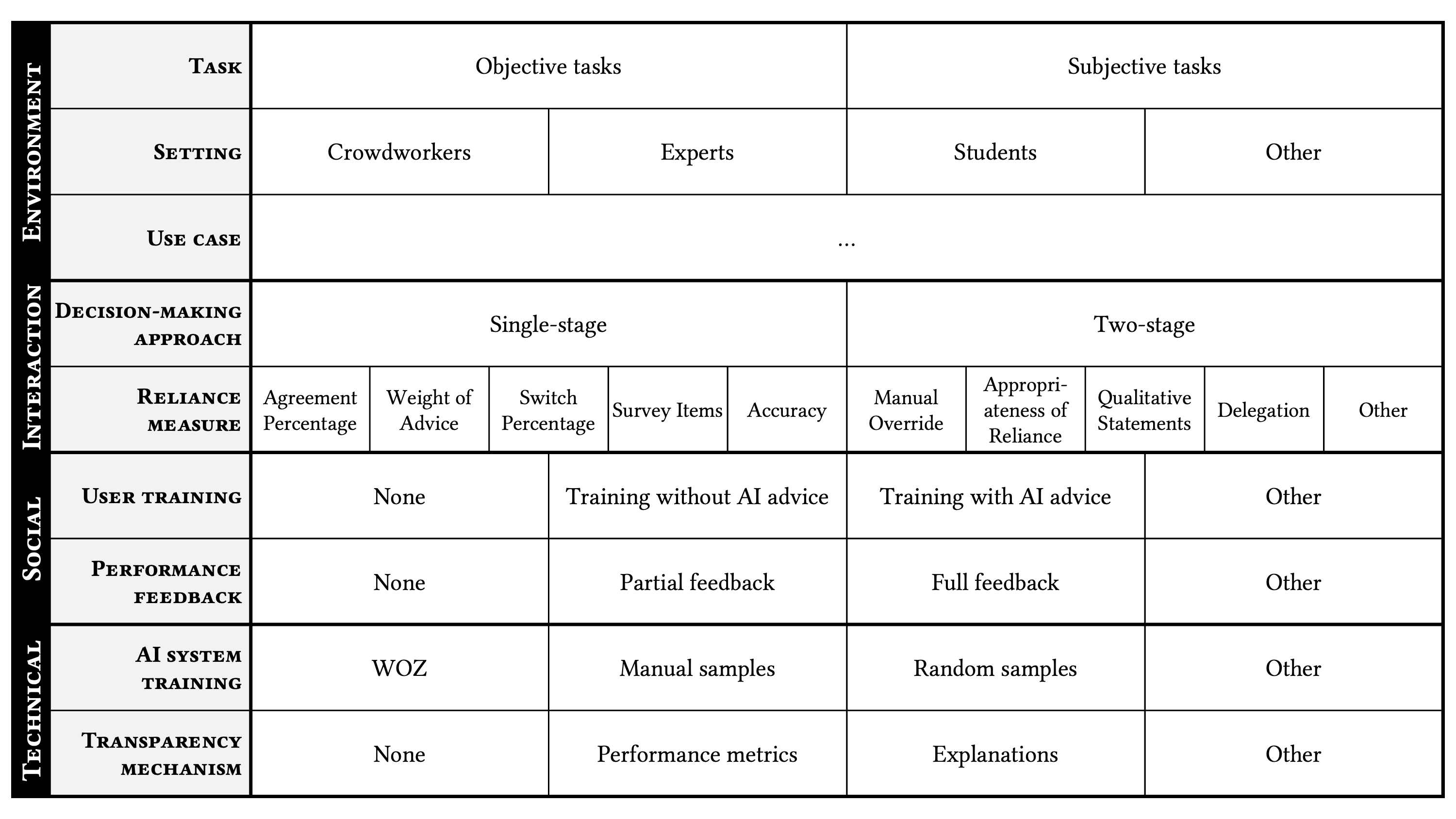}
    \caption{Morphological box for the design of AI reliance studies.}
    \Description{The figure illustrates a morphological box for the design of AI reliance studies. On the left are the categories of the classification framework. On the right are the individual attributes identified in the literature that collectively summarize the design of AI reliance studies.}
    \label{fig:morph_box}
\end{figure}

A review of the extant literature on AI reliance reveals that the articles exhibit a number of common patterns, which can be grouped according to the concepts presented in \Cref{tab:concepts} on page \pageref{tab:concepts}. It also became evident that the requisite information could at times not be extracted with ease, as some articles lacked clarity. As previously indicated, user training was a notable example. We invite all studies on AI reliance to use the concepts introduced in \Cref{tab:concepts} on page \pageref{tab:concepts} as a basis for presenting the required information. 

To facilitate the work of researchers, we have transformed this table into a morphological box, presented in \Cref{fig:morph_box}. 
The attributes for the morphological box are based on the findings concerning general patterns in the literature and presentations in the results section. The morphological box can serve as guidance for future studies on AI reliance. It is not our intention to assert that some approaches or attributes are inherently superior; this morphological box can, however, serve as a reference for the design of studies. It is also recommended that researchers provide a rationale for the selection of specific approaches. For instance, future studies on AI reliance should clearly justify the choice of a single-stage over a two-stage decision-making approach or the decision to have user training. By doing so, we can establish a unified framework for AI reliance studies and provide guidance to researchers in the design of their studies. Ultimately, this leads to a common understanding and comparable results of AI reliance researchers. 

\subsubsection*{Retrospective Application of the Morphological Box}
In addition to using this morphological box as guidance, research can also be conducted with the objective of expanding it. For that, a retrospective perspective on current research is gained by applying the morphological box. This review represents an initial and indispensable step toward a unified and guiding framework for AI reliance researchers. The morphological box in reference \Cref{fig:morph_box} just reflects the current state of research. We demonstrate its applicability to two exemplary research projects with researchers outside the author team \cite{morrison2023evaluating,holstein2023toward} and document the results in \Cref{sec:exemplaryapplication}. Future research may present extensions to it, either by introducing new attributes for the individual subconcepts or by introducing new subconcepts altogether. These may, for instance, include the underlying interaction structure between user(s) and AI, should more than one user be considered. Before extending the morphological box, further substantial research on the emerging issues is recommended. 

\subsubsection*{Prospective Application of the Morphological Box}
While our review offers a comprehensive overview of AI reliance research, the current literature still reveals important gaps, particularly concerning the systematic investigation of the core influencing factors: human characteristics, task characteristics, and AI system features (e.g., explainability). Although these elements are addressed individually across studies, there remains a lack of integration and consistency in how they are both conceptualized and measured. Our morphological box provides a structured foundation to address this fragmentation and can serve as a guiding tool for future research. 

\subsubsection*{Exemplary research designs} To support researchers, we propose three exemplary research designs.  

\textit{Research Design 1: Measuring Switch Behavior in Experts:} Current research often relies on indirect or subjective measures (e.g., surveys, interviews) to assess expert reliance, but rarely uses switch percentage to capture actual behavioral change. This limits comparability and interpretability, especially in high-stakes expert contexts. The morphological box highlights the potential of using switch percentage to better assess whether experts truly revise their decisions in response to AI advice.

\textit{Research Design 2: Comparing One-Stage vs. Two-Stage Designs:} A direct comparison of one-stage and two-stage designs under otherwise identical conditions would allow researchers to isolate the effects of anchoring and identify differences in measured reliance. The morphological box helps clarify where this design decision matters and encourages systematic experimentation across conditions.

\textit{Research Design 3: Training in Subjective Tasks:} The morphological box reveals a current research gap: studies on subjective tasks (e.g., music or film preferences) rarely incorporate user training, often assuming calibration is irrelevant. Yet especially in such tasks, users need experience to judge whether the AI aligns with their personal values or preferences. Future studies could fill this gap by systematically investigating how training influences reliance in subjective, preference-driven domains.

In the digital appendix, we present these three exemplary research paths through the morphological box, illustrating how different factor combinations can be systematically explored. We encourage future work to use this tool not only for structuring study designs but also for ensuring comparability and cumulative knowledge building.

\subsubsection*{Recommendations for researchers} For research on AI reliance, we derive the following methodological recommendations for future studies on AI reliance. 
\begin{itemize}
    \item Be aware of the individual pros and cons of using one-stage vs. two-stage decision designs. 
    \item Incorporate user training and/or performance feedback when studying reliance dynamics or calibration effects.
    \item Tailor reliance measures to task types: use switch rates or weight of advice in objective tasks; use delegation rates, click behavior, or qualitative markers in subjective contexts.
    \item Explicitly report baseline human accuracy and task complexity to contextualize reliance measures.
\end{itemize}

In summary, despite the growing body of research on AI reliance, our analysis reveals that many studies focus narrowly on isolated factors, such as task complexity \cite{cabrera_improving_2023} or explanation style \cite{schemmer_appropriate_2023}, without systematically considering the broader interaction between human characteristics (e.g., domain knowledge, cognitive load) \cite{keding_managerial_2021}, task characteristics (e.g., decision risk, subjectivity) \cite{narayanan_how_2023}, and AI system features (e.g., transparency, uncertainty information) \cite{jakubik2022empirical}. Few studies examine these dimensions in combination or track how they affect reliance behaviors across settings. This fragmentation hinders theory development and the accumulation of generalizable insights. As a response, we propose the morphological box, hoping to enable researchers to design and compare studies more systematically by mapping existing research and highlighting underexplored intersections across human, task, and system dimensions.

\subsection{A Research Agenda on AI Reliance}

Overall, we identify the following six directions for future and more in-depth research on AI reliance: (1) improving the external validity of studies (2) increasing the reliability of AI reliance measurements; (3) understanding the antecedents of AI reliance; (4) Understanding AI reliance over time;  (5) understanding reliance of more than one user, (6) revisiting AI reliance in the light of generative AI.

\subsubsection{Improving the External Validity of Studies}
The results show that reliance is considered in a multitude of use cases, ranging from autonomous vehicle control \cite[e.g.][]{srivastava_improving_2022} to house price predictions \cite[e.g,][]{chiang_youd_2021} or solving ethical concerns \cite[e.g.][]{tolmeijer_capable_2022}. However, these use cases are most often considered in controlled experiments, where WOZ approaches or manually picked sets of decision tasks dominate the interaction with the AI system. Further, most experiments are conducted with crowdworkers of MTurk or Prolific. This relates to the previously mentioned \textit{precision-realism tension}. On the one hand, researchers are drawn to cases with high real-world relevance, often incorporating real-world data into their systems or experiments. On the other hand, they employ experimental methods that potentially undermine the connection to the real world. While such experimental approaches have the positive effect of easy and fast data collection, they come with the downside of missing external validity, as real AI systems with real users are rarely considered.

The perception of randomness among humans is subject to bias. For instance, when asked to generate a random series, humans tend to generate series with higher-than-expected alternation rates. This is because humans tend to perceive clumps or streaks as not truly random, while they actually are \citet{bar1991perception}. Conversely, AI is susceptible to randomness due to its probabilistic nature. Each time a study presents a manually selected set of decision instances or a WOZ study, users are confronted with a human bias from the outset. While numerous articles attempt to present a representative set of decision instances, they do not deliver any confirmation of the true randomness. Consequently, we call for more research where users (decision-makers) of AI systems are confronted with genuine and randomly selected AI output.
 
An illustrative example of a phenomenon that occurs in real-world AI interactions is the presence of outliers. They are an inherent and integral part of any AI system, and are often targeted to be detected \cite{boukerche2020outlier}. While many approaches aim to minimize their occurrence, there is always the possibility of outliers occurring due to the probabilistic nature of current AI approaches. Examining the reliance on AI systems after a user is confronted with an outlier is therefore also important, and is, for example, done by \citet{leffrang_people_2023}. Thus, we specifically call for more research where he users (decision-makers) of AI systems encounter naturally occurring outliers in AI output. 

A further limitation of current research is that many studies use crowdworkers rather than actual end-users or decision-makers interacting with the system for its intended purpose. The use of crowdworkers, such as MTurk users, has been questioned and may render low data quality \citet[e.g.][]{hsueh-etal-2009-data, kennedy2020shape, chmielewski2020mturk}. While a substantial body of research is dedicated to enhancing data quality and providing best practices \citet[e.g.][]{kees2017analysis}, concerns about the external validity and generalizability of the results persist. A recent discussion has emerged regarding the potential for crowdworkers to employ AI systems, such as LLMs, to automate their tasks \cite{kauffman_turk_2023}. This would also diminish the reliability of using crowdworkers, as the responses are not human-generated but rather AI-generated, removing the human element from the equation. This would result in investigations of the reliance of AI systems on other AI systems. We therefore call for more research with more natural and controlled users. 

Another aspect related to crowdworkers or random sampling of large participant groups is the \textit{subjectivity dilution}, where questions that rely on informed guesses or predictions are considered subjective assessments. This assumption aligns with the use of crowdworkers or random samples, as everyone can be said to have an opinion or preference. However, sometimes the questions asked extend beyond individual, private opinions and may require logical assessment, consideration of facts, or similar. Based on \citet{kahneman2011thinking}, we speculate that real-world domain experts accustomed to making such assessments frequently and aware of their cognitive biases may develop a different level of confidence in their evaluations compared to those who simply make random guesses when confronted with estimation questions, such as those about preferences of a population. This, in turn, will significantly impact reliance behavior. We therefore call for more research involving true experts. 

It becomes evident that real users and real environments are rarely studied. This also raises the question of high-stakes decision-making tasks. While several of the examined use cases may be considered high-stakes decisions, such as medical decision-making \cite[e.g.][]{du_role_2022,panigutti_understanding_2022,wysocki_assessing_2023}, the consequences of these decisions are rarely observed. For instance, in the case of kidney donors \cite{narayanan_how_2023}, no actual patient is affected by the decision; therefore, the decision-maker may approach the problem differently if they were aware that a real patient's life or death depended on the outcome. Consequently, we call for more research addressing the lack of high-stakes decisions.

Overall, the current research is subject to \textit{external validity deficit} as mentioned in \ref{sec:settings}, suggesting that the decision-making scenarios in the studies may not accurately reflect real-world decision-making. We therefore call for research that examines more real-world settings and existing AI systems. There are numerous existing AI systems that could be used to investigate reliance. For instance, researchers with access to Netflix or Amazon data could investigate user reliance when faced with their recommendation algorithms. While there is undoubtedly a considerable amount of research on these algorithms \cite{amatriain2015recommender, hallinan2016recommended,linden2003amazon}, it seems to be concerned with the accuracy and design of the AI systems themselves rather than user reliance. While reliance is related to accuracy, it is not the same. To illustrate, if the Netflix algorithm were to predict a movie recommendation with 100\% accuracy based on the training data, this performance could only be achieved in real life if the user relies on these recommendations. This reliance has more facets than the pure performance of the AI advice, which is highlighted by the sociotechnical perspective. Consequently, it is not only the performance that must be evaluated but also the extent of AI reliance, particularly in real-world settings. 

\subsubsection{Increasing the Reliability of AI Reliance Measurements}
One potential explanation for the lack of external validity in the current literature may be the difficulty of measuring reliance as indicated by the observation on \textit{measures discrepancy}. In particular, in real-world settings, it is unclear whether a human decided to do something because of the AI or whether the human simply agreed with the AI, but the AI did not influence their decision-making. For example, if a consumer purchases a product after it has been recommended by Amazon's algorithm, it is challenging to ascertain whether the consumer would have made the purchase regardless of the recommendation. Conversely, the question of whether a human decision-maker was influenced by AI advice becomes more straightforward in experiments where researchers can control the environment. 

In controlled experiments, research sometimes employs a two-stage decision-making approach where the user must first make their own decision without any AI support. They get AI advice only afterwards and use it to change their decision. A change can then be attributed to the AI advice and be defined as reliance. In contrast, the single-stage decision approach directly confronts the user with the AI advice, rendering any such consideration impossible. This two-stage decision-making approach offers several advantages. It is designed to counteract some human cognitive biases, such as anchoring or priming, which occur by presenting the AI advice first. Additionally, research in the field of advice-taking indicates that this approach can lead to enhanced performance overall  \citet{sniezek1995cueing}. The two-step approach allows for the identification of instances where a user may have agreed with the AI advice but would have reached the same conclusion independently. This approach also has a potential drawback, as humans may be reluctant to alter their initial decisions. One indication of status quo bias is that humans are cautious about changing the status quo \cite{samuelson1988status}. This is evidenced by their inclination to remain at the current status, even when presented with new information. Furthermore, the initial answer could be seen as self-anchoring, with the individual's initial response becoming a reference point for subsequent decisions.

\begin{figure}[htbp]
    \centering
    \includegraphics[width = 0.8\textwidth]{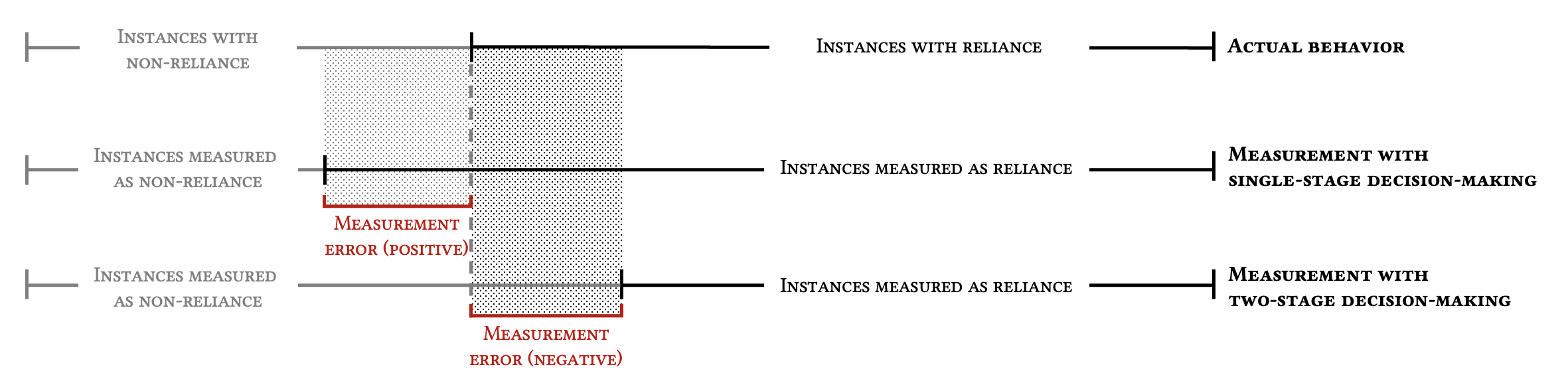}
    \caption{Tradeoff between metrics for single-stage and two-stage decision-making processes. Metrics for single-stage decision-making processes might also count instances where the user's decision and the AI's advice coincide, but the user does not rely on the AI system, but arrives at the decision on their own, leading to (positive) measurement error. Metrics for two-stage decision-making processes might not include instances where the user has already arrived at the AI advice before the AI advice, but would have relied on the AI advice if it had been different, leading to (negative) measurement error.  Note: Boxes are not true to size and are for illustration purposes only.}
    \Description{The figure illustrates the tradeoff between metrics for single-stage and two-stage decision-making processes, as represented by tree lines. The top line presents an underlying hypothetical ground truth of instances without reliance and instances with reliance. The second line represents the measurement with single-stage decision-making and indicates the potential for a measurement error in classifying instances without reliance as reliance. The third line represents the measurement with two-stage decision-making and suggests the possibility of a measurement error in not classifying instances with reliance on such.}
    \label{fig:1-two-stage}
\end{figure}

Both approaches use a multitude of metrics, as illustrated in \Cref{fig:measures} on page \pageref{fig:measures}. While some of these metrics are directly inferred from either approach, such as agreement percentage and switch percentage, others are only loosely connected, such as survey items and qualitative statements. In particular, the latter metrics use self-reported data and are therefore suboptimal for collecting behavioral data. Consequently, we view them as supplementary rather than standalone metrics. Concerning the single-stage and two-stage decision-making approaches, analogies can be drawn between the concepts of precision and recall. While the single-stage decision-making approach increases recall of instances where the human relies on the AI system, the two-stage decision-making approach increases precision. These potential errors are illustrated in \Cref{fig:1-two-stage}. It is currently not possible to quantify these measurement errors; consequently, it is not possible to determine which approach is superior. The choice between the two approaches depends on the specific errors that one wishes to avoid. The decision to select either one must therefore be made with careful consideration. 

In summary, we see that both the single-stage and two-stage decision-making approaches have advantages and disadvantages. The single-stage decision-making approach is easy to measure, straightforward, and understandable. Reliance is, however, overestimated due to the classification of numerous instances as such, despite the user ultimately being the decision-maker. The two-stage decision-making approach is more robust; therefore, reliance can be identified with greater precision. Instances where a user makes a different decision because they have to make a prior decision may, however, be overlooked. We call for more research on how to control the biases of both approaches and on guidance on when to apply which approach (also see research direction 2 in \Cref{fig:morph_boxapplicationfuture}).  

\subsubsection{Understanding the Antecedents of AI Reliance}
The sociotechnical perspective on AI reliance identifies four distinct angles through which influences on AI reliance can be identified: the social component, the technical component, the interaction between these two, and the environment. All these factors are antecedents of AI reliance for decision-making. Only with this comprehensive perspective can we fully acknowledge the influences on AI reliance. A review of the literature reveals that most articles only consider individual components of the STS. Even within these components, there is often no consensus on the factors that influence reliance on AI systems. As indicated with regard to \textit{multi-factorial effects neglect}, what is notably lacking is a clear understanding of how different factors interact with one another as antecedents to reliance behaviors, and which of these factors are especially crucial in fostering appropriate reliance. In the following section, we discuss current approaches in the various components of STSs. In conclusion, we urge further research to obtain a more comprehensive view of the antecedents of AI reliance. 

A significant proportion of articles in this field focus on the technical component---the AI system itself---and its influence on reliance. The design aspects of AI systems are frequently the primary considerations, with explanations being a key area of interest. It is also evident in the literature that transparency mechanisms, such as explanations, do not have a monotonic influence, and explanations do not always lead to the same results \cite{schemmer2022meta}. Consequently, an exclusive focus on explanations as indicated by the observation of \textit{explainability fixation}, or the design of the technical component in general, is insufficient for a comprehensive understanding of AI reliance. It is imperative to consider the users interacting with the AI system. 

The social component, referring to the human decision-maker, can also influence AI reliance. Humans are prone to cognitive biases \cite{kahneman2011thinking}, and AI systems have been shown to induce various such biases, most notably the anchoring or automation bias \cite{rastogi_deciding_2022}. The effect of the bias may also depend on humans themselves, with some people being more prone to cognitive biases than others. For instance, domain knowledge has been shown to reduce these cognitive biases \cite{krems1994experts}. This may lead to higher self-reliance and, consequently, less reliance on AI systems. Some articles examine the effects of domain knowledge  \citet[e.g.][]{bayer_role_2022}. Consequently, the human itself, in terms of characteristics such as being experts or novices, may influence the level of reliance. We call for more research on how the human component influences AI reliance.
 
The environment in which humans and AI interact could also influence reliance. Some articles suggest that time pressure is an influence \cite{cao_how_2023}. \citet{barr_kumarakulasinghe_evaluating_2020} point out that physicians' active work environment might also influence their reliance. To illustrate, a physician with the same computer vision system may exhibit excessive reliance on the system when a patient's condition is critical and a decision must be made in seconds. However, in an environment where the physician has ample time to make decisions, there might be appropriate reliance on the AI system. The environment in which humans and AI interact may therefore influence the degree of AI reliance. 

Finally, the interaction itself can influence reliance. A review of the literature reveals that there are two main approaches to decision-making and subsequently measuring reliance: single-stage and two-stage. Both influence the interaction and subsequently AI reliance. A single-stage decision-making approach facilitates fast decision-making and quick acceptance of the AI advice. In contrast, a two-stage decision-making approach enforces slow thinking, and the decision-maker must invest more effort in coming up with a decision. Cognitive forcing has been identified as a factor that influences reliance \cite{bucinca_trust_2021}. The interaction itself, therefore, plays a role in determining the extent to which humans rely on AI advice. 

The influences on reliance can be conceptualized as a multidimensional space. For instance, a user interacting with the same AI system may exhibit varying degrees of reliance based on environmental factors, such as time constraints. Alternatively, a user interacting with the same AI system in the same environment may exhibit varying degrees of reliance based on the decision-making approach. This is because the same user may have a different reliance behavior when confronted with a single-stage decision-making process as opposed to a multistage approach. It is evident that current research articles tend to focus on a single (or at most two) dimension of AI reliance. A more comprehensive approach is, however, required to fully understand the nature of AI reliance. Specifically, there is a lack of understanding of how various dimensions or individual influence factors interact with each other with regard to reliance. For instance, it is likely that time pressure can interact with the provision of explanations and their form. We therefore call for more research on all dimensions. This would, however, lead to virtually unlimited combinations that would require attention. Nonetheless, we claim that focusing on combinations that occur frequently in the real world can help select appropriate combinations, while ensuring external validity.  

\subsubsection{Understanding AI reliance over time}

\begin{figure}[!htbp]
    \centering
    \includegraphics[width= 0.6\textwidth]{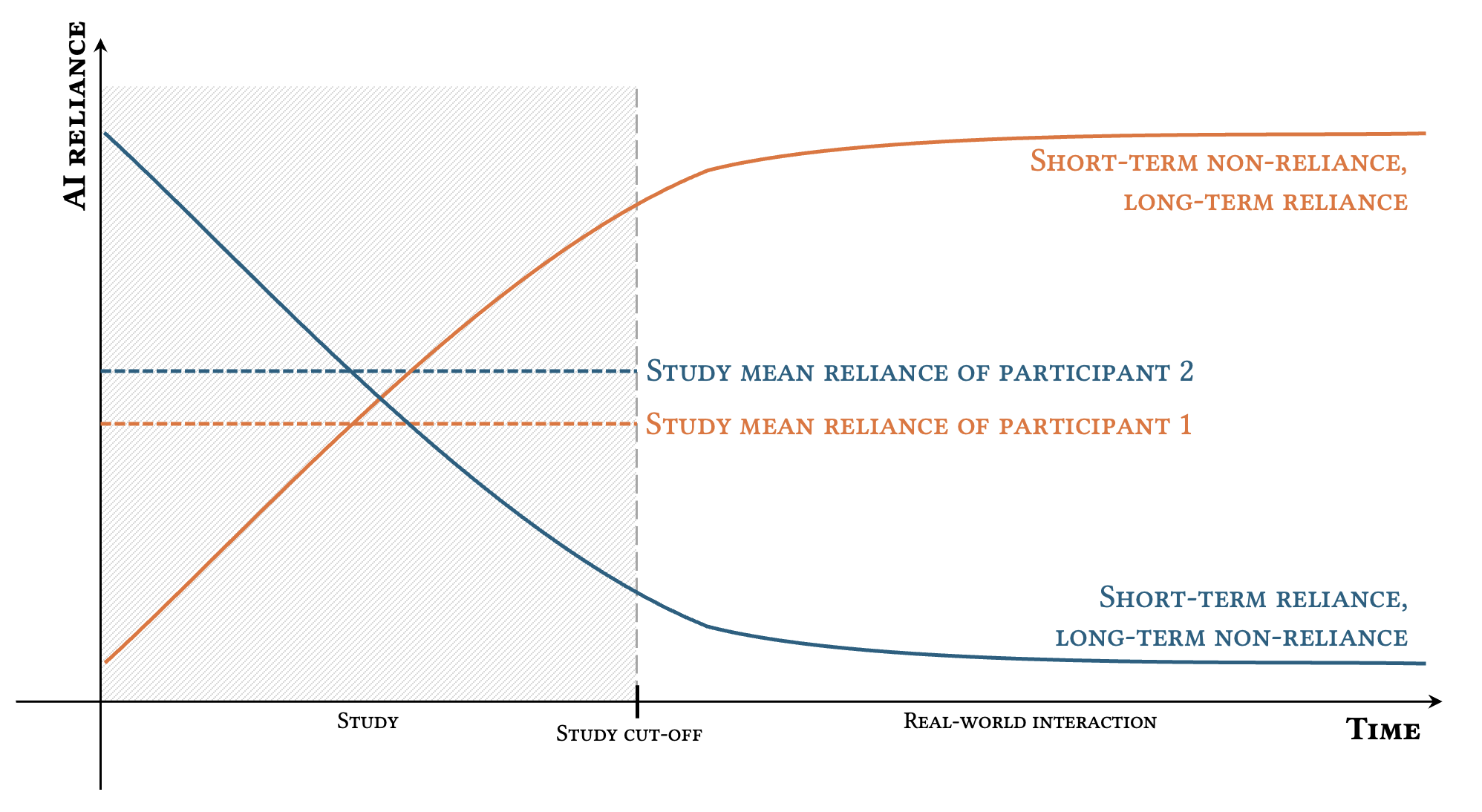}
    \caption{Possible reliance change over time, where results unaware of the change over time might lead to misleading conclusions.}
    \Description{The figure depicts an axis diagram with the x-axis representing time and the y-axis indicating the level of AI reliance. The orange graph, labeled "short-term non-reliance, long-term reliance," exhibits a step incline and ultimately reaches a high level. A blue graph, labeled "short-term reliance, long-term non-reliance," exhibits a step decline and ultimately reaches a low level. Additionally, a vertical line is positioned near the knee of both graphs, representing a study cut-off. Two dashed lines, colored according to the respective graphs, represent the mean of the left side of this study's cutoff.}
    \label{fig:relianceTime}
\end{figure}

Most articles investigate users interacting with an AI system on multiple occasions, yet within a short time frame of an experiment, as described by the \textit{temporal effects neglect}. The reliance level (e.g., agreement fraction) is then calculated by averaging all decision instances. For instance, if a user relies on the AI system eight out of ten times, the reliance level is 0.8. Alternatively, if the WOA is calculated, the reliance is defined as the mean WOA over these ten decision instances. This average reliance consideration neglects potential trends of reliance behavior. When a user interacts with an AI system on multiple occasions, it is possible that the effect may only become apparent over time. However, only a few articles consider the changes to reliance over multiple interactions, for example \citet{leffrang_people_2023} focused on the effect of an outlier and how reliance changes after this outlier. 

To illustrate this point, consider a study that investigates reliance on AI advice and concludes that users, in general, have a moderate level of reliance. A longitudinal field trial may, however, reveal that after some time, some users start to always rely on AI advice, while others stop. Consequently, the findings of the study may become invalid after a certain period of time. An extreme case of this phenomenon is illustrated in \Cref{fig:relianceTime}. In \Cref{fig:relianceTime}, the (hypothetical) participant 1 (orange) increases the reliance over the time of the study and over time will always rely on the AI advice, which we denote as \textit{short-term non-reliance, long-term reliance}. The (hypothetical) participant 2 (blue) in \Cref{fig:relianceTime} is the opposite, which we denote as \textit{short-term reliance, long-term non-reliance}. While both will have similar reliance behavior in a study, both users have fundamentally different reliance over a longer period of time. A study that only measures average reliance will not detect this decline. This also has implications for the real-life use of AI systems. For example, if a new AI system is tested and the study concludes that reliance is high, in practice, the same tool might not impact the real world if it follows a similar pattern as (hypothetical) participant 2 in \Cref{fig:relianceTime}. Consequently, we posit that studies should be guided by a conscious decision as to whether to consider the reliance change over time or to limit their scope to the average reliance level. 
 
Overall, it is important to consider the concept of reliance over time. Over time, there can be a circular relationship between humans and reliance as both adapt to prior outcomes. There is a clear need for further research into this phenomenon of reliance and users changing over time.

\subsubsection{Understanding reliance of more than one user}

In the context of research on human-AI teams, most studies focus on a single user and an AI system. Similarly, research on the performance of human-AI teams tends to examine the combined output of a single user and AI system, with the expectation that this team will outperform the individual components. Unsurprisingly, the literature on AI reliance mirrors this individualistic orientation. To our knowledge, only one study explicitly investigates reliance dynamics in group decision-making, using the COMPAS algorithm as a case \citet{chiang_are_2023}. 

In real-world settings, we posit that most cases involve multiple users or even multiple parties with conflicting interests. For instance, numerous articles address house price predictions with a focus on a single user \cite{chiang_youd_2021, chiang_exploring_2022, prabhudesai_understanding_2023}. The impact of such AI systems on real-life settings is contingent upon the degree of reliance placed upon them by both the seller and the buyer of houses. More precisely, in practice, both buyer and seller interact with the system---and rely on it in different, possibly conflicting ways. Given the inherent conflict of interest between these two parties, further research is warranted to investigate the extent to which these systems can be relied upon. 

In addition to the existing literature on AI reliance, recent studies investigate the integration of AI in creative, collaborative settings \cite{shin2023integrating}. The objective of the AI system is to support the ideation of multiple users. These settings are, however, all concerned with collaborative settings, where multiple users come together with common goals and interests, excluding settings with conflicting interests. Looking forward to guiding future research more concretely, we recommend that studies on multi-user reliance use controlled lab experiments where user roles and coordination mechanisms are explicitly manipulated (e.g., shared vs. individual decision authority). A two-stage decision-making setup can be used to measure role-specific shifts in reliance.

In conclusion, it is evident that there is a necessity to consider settings with multiple users and parties that are able to capture real human interaction. This can be achieved by extending the currently considered use cases to encompass more stakeholders. As previously discussed, AI systems supporting house price predictions must be relied upon by both the buyer and seller. Similarly, AI systems supporting doctors in diagnoses must also be relied upon by the patients in order to have a real-life impact. 

\subsubsection{Revisiting AI reliance in the light of Generative AI}

Generative AI has introduced a fundamentally different class of systems into the AI reliance discourse---systems that not only provide answers but co-create content with users. Unlike traditional AI systems that support bounded decisions with clear correctness criteria, generative models often operate in subjective, open-ended domains (e.g., creative writing, image generation, coding), where correctness is ill-defined and outcomes emerge through iterative co-creation. In such tasks, users may only have a vague intention or aesthetic preference, and the AI system contributes to shaping---not just executing---the output. This challenges the idea of a distinct human decision point prior to AI advice. Consequently, the forthcoming generative AI systems will form an additional discourse to AI reliance research. 

It is noteworthy that some of the literature on AI reliance already employs generative AI models, most notably large language models, as evidenced by the work of \citet{schmitt_towards_2021} and \citet{goyal-etal-2023-else}. These approaches use large language models as the underlying foundation, yet the output remains a closed form of advice, such as an answer to a specific question, as opposed to a creative and generative task. 

The most prevalent example of generative advice currently is GPT models, which are accessible through the interface of ChatGPT \cite{Chatgpt2023}. The advent of generative AI systems has enabled everyday users to use these systems, but it is not yet clear how they are integrated into users' decision-making processes. For instance, if a system preformulates a response to an email and the human user adapts some passages, it is unclear what reliance really means in that case. In the context of text generation, the text similarity between the LLM-generated output and the human text might be used as a quantification. This also raises the question of text similarity, as computer science has developed numerous metrics at different levels of granularity, including character-level, word-level, and even semantic similarity. 

As these systems are increasingly used, there is a pressing need to conceptualize and define reliance on these systems. Once this conceptualization is established, research should aim to identify methods for quantifying reliance on generative AI output. Future studies should assess reliance using behavioral proxies such as suggestion acceptance rates, degree of content modification, or response latency, rather than binary agreement measures. Experimental tasks might include creative writing or data storytelling, where objective correctness is not well-defined.

\subsection{Practical Guidelines for Practitioners}
While this review is primarily oriented toward the academic study of AI reliance, it also provides several actionable insights for practitioners involved in the development, deployment, and evaluation of AI systems. System designers should aim not for maximum user trust, but for calibrated reliance that reflects the strengths and limitations of the AI system. This means designing interfaces and outputs in ways that support informed user judgment, particularly in tasks involving uncertainty or high stakes. Mechanisms such as explanations, uncertainty cues, or feedback loops can be valuable here, especially when deployed in combination and tailored to user needs. Further, practitioners need to assess and support reliance in context. The appropriateness of reliance depends on several interdependent factors---such as the objectivity of the task, the experience and expertise of the user, and the extent to which users are trained to understand and evaluate the system's output. Evaluations should consider these contextual elements explicitly rather than relying on one-size-fits-all performance metrics. Also, AI systems should be designed to support dynamic calibration over time. Users' reliance behaviors may shift as they gain experience with a system or encounter feedback on its performance. Where feasible, systems should provide opportunities for users to receive feedback---either through explicit outcome validation or indirect cues that help recalibrate expectations. In multi-user settings, practitioners must consider how reliance plays out across different roles and interaction patterns. For instance, systems that influence decisions involving multiple stakeholders---such as clinicians and patients, or buyers and sellers---require careful design to ensure that responsibility, access to AI advice, and interpretation of outputs are distributed in ways that align with the intended decision process. Finally, the morphological box can serve as a practical planning tool. Practitioners can use it to map out key design dimensions---such as user characteristics, task types, AI features, and feedback mechanisms---when developing new systems or planning evaluations. By systematically reflecting on how different configurations may influence user reliance, the box can help anticipate risks of over- or underreliance and guide the implementation of safeguards or interventions.

\section{Conclusion}
\label{sec:conclusion}
In this study, we conduct a survey of existing literature on AI reliance. We employ a sociotechnical perspective, as this allows us to gain a comprehensive understanding. We derive concepts based on the four components of an STS to classify AI reliance literature and to classify current literature. Furthermore, we discuss current issues and topics related to AI reliance. This review aims to support future researchers on AI reliance. We provide researchers with a framework to evaluate their AI systems, enabling them to classify and present their results. We also identify future avenues for research on AI reliance in the form of a research agenda, assisting researchers in their endeavors. 

\bibliographystyle{ACM-Reference-Format}
\bibliography{bib-files/bib}

\section*{Appendix}
\appendix
\section{Database Queries}

Query Execution Date: 2024-01-25

\paragraph{\textbf{ACM DL}}:

\begin{itemize}
    \item[] \textbf{Filter} \begin{itemize}
    \item year: 2010 - 2023
\end{itemize}
    \item[] \textbf{Query}
    \begin{itemize}
    \item[] \texttt{(Abstract:("Artificial Intelligence" OR "AI" OR "Machine Learning" OR "ML" OR "Deep Learning" OR "DL") OR Title:("Artificial Intelligence" OR "AI" OR "Machine Learning" OR "ML" OR "Deep Learning" OR "DL") OR Keyword:("Artificial Intelligence" OR "AI" OR "Machine Learning" OR "ML" OR "Deep Learning" OR "DL")) AND (Abstract:("Reliance" OR "Underreliance" OR "Overreliance") OR Title:( "Reliance" OR "Underreliance" OR "Overreliance") OR Keyword:( "Reliance" OR "Underreliance" OR "Overreliance"))}
    \end{itemize}
\end{itemize}

\paragraph{\textbf{AISeL}}:

\begin{itemize}
    \item[] \textbf{Filter} \begin{itemize}
        \item AISeL does not allow for easy filtering, the year constraint has been applied after downloading all papers
\end{itemize}
    \item[] \textbf{Query}
    \begin{itemize}
    \item[] \texttt{(abstract:("Artificial Intelligence" OR  "AI" OR  "Machine Learning" OR  "ML" OR  "Deep Learning" OR  "DL") OR  title:("Artificial Intelligence" OR  "AI" OR  "Machine Learning" OR  "ML" OR  "Deep Learning" OR  "DL")) AND (abstract:("Reliance" OR  "Overreliance" OR  "Underreliance") OR  title:("Reliance" OR  "Overreliance" OR  "Underreliance"))}
    \end{itemize}
\end{itemize}

\paragraph{\textbf{SCOPUS}}:

\begin{itemize}
    \item[] \textbf{Filter} \begin{itemize}
        \item Top 25\% journals according to Citescore on \url{https://www.scopus.com/sources}
        \item Year: 2010-2023
        \item Document type: ``Conference paper'', ``Article ''
\end{itemize}
    \item[] \textbf{Query}
    \begin{itemize}
    \item[] \texttt{TITLE-ABS ( ( "Artificial Intelligence" OR "AI" ) AND ( "Reliance" OR "Overreliance" OR "Underreliance" ) ) OR ( TITLE-ABS ( "Artificial Intelligence" OR "AI" ) AND AUTHKEY ( "Reliance" OR "Overreliance" OR "Underreliance" ) ) OR ( TITLE-ABS ( "Reliance" OR "Overreliance" OR "Underreliance" ) AND AUTHKEY ( "Artificial Intelligence" OR "AI" ) ) }
    \end{itemize}
\end{itemize}

\paragraph{\textbf{IEEE}}:

\begin{itemize}
    \item[] \textbf{Filter} \begin{itemize}
        \item Year: 2010-2023
        \item Document type: ``Conferences'' or ``Journal''
\end{itemize}
    \item[] \textbf{Query}
    \begin{itemize}
    \item[] \texttt{((("Document Title" : "Artificial Intelligence" OR "Document Title" : "AI" OR "Document Title" : "Machine Learning" OR "Document Title" : "ML" OR "Document Title" : "Deep Learning" OR "Document Title" : "DL" ) OR ("Abstract" : "Artificial Intelligence" OR "Abstract" : "AI" OR "Abstract" : "Machine Learning" OR "Abstract" : "ML" OR "Abstract" : "Deep Learning" OR "Abstract" : "DL") OR ("Author Keywords" : "Artificial Intelligence" OR "Author Keywords" : "AI" OR "Author Keywords" : "Machine Learning" OR "Author Keywords" : "ML" OR "Author Keywords" : "Deep Learning" OR "Author Keywords" : "DL")) AND (("Document Title" : "Reliance" OR "Document Title" : "Overreliance" OR "Document Title" : "Underreliance") OR ("Abstract" : "Reliance" OR "Abstract" : "Overreliance" OR "Abstract" : "Underreliance") OR ("Author Keywords" : "Reliance" OR "Author Keywords" : "Overreliance" OR "Author Keywords" : "Underreliance")))}
    \end{itemize}
\end{itemize}

\newpage
\section{Applications of the morphological box}
\label{sec:exemplaryapplication}

\begin{figure}[htbp]
    \centering
    \includegraphics[width = 0.7\textwidth]{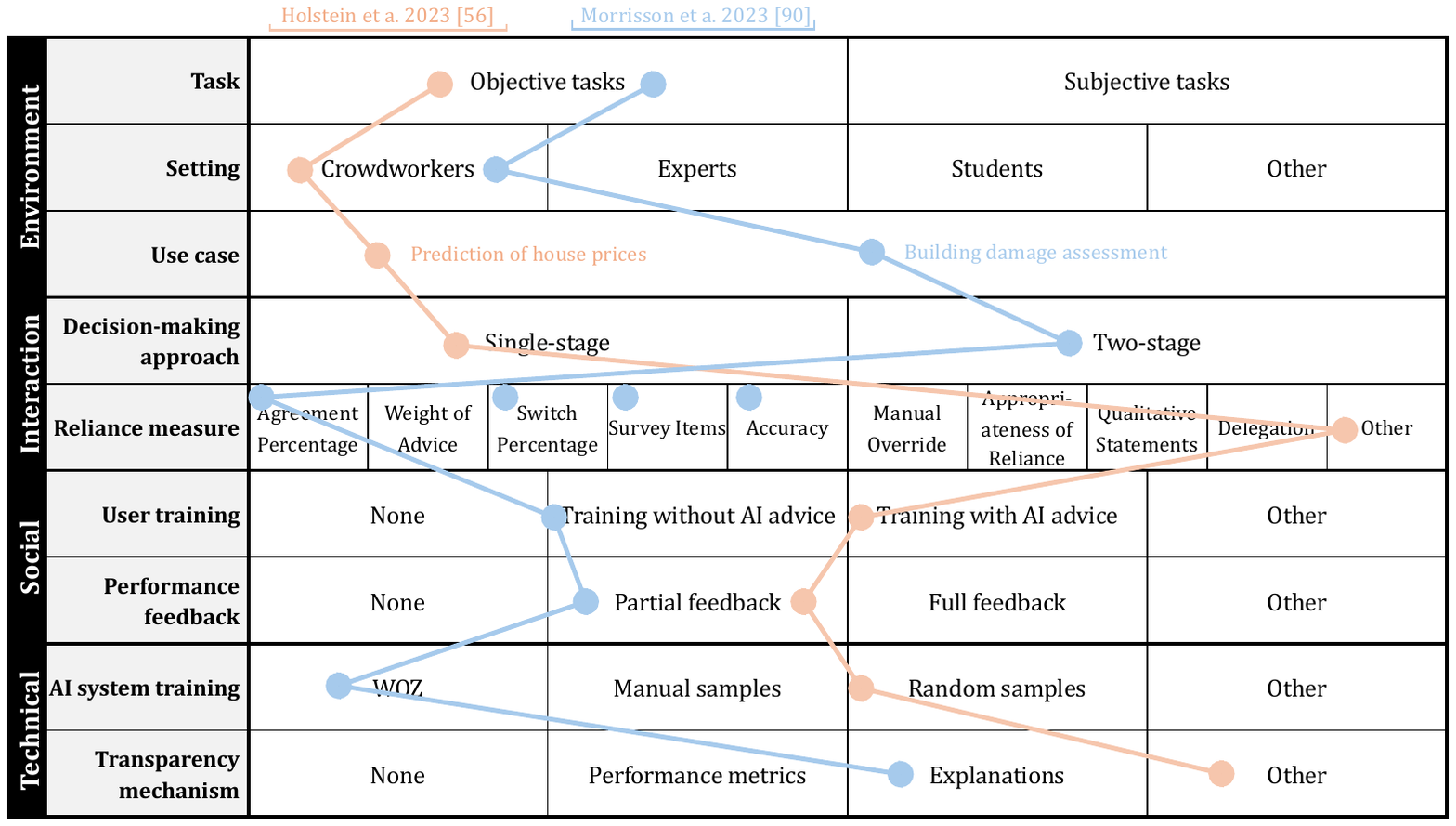}
    \caption{Exemplary application of the morphological box to the existing research articles of \citeauthor{morrison2023evaluating}\cite{morrison2023evaluating} and \citeauthor{holstein2023toward}\cite{holstein2023toward}.}
    \Description{The figure illustrates the application of the morphological box.}
    \label{fig:morph_boxapplication}
\end{figure}

\begin{figure}[htbp]
    \centering
    \includegraphics[width = 0.7\textwidth]{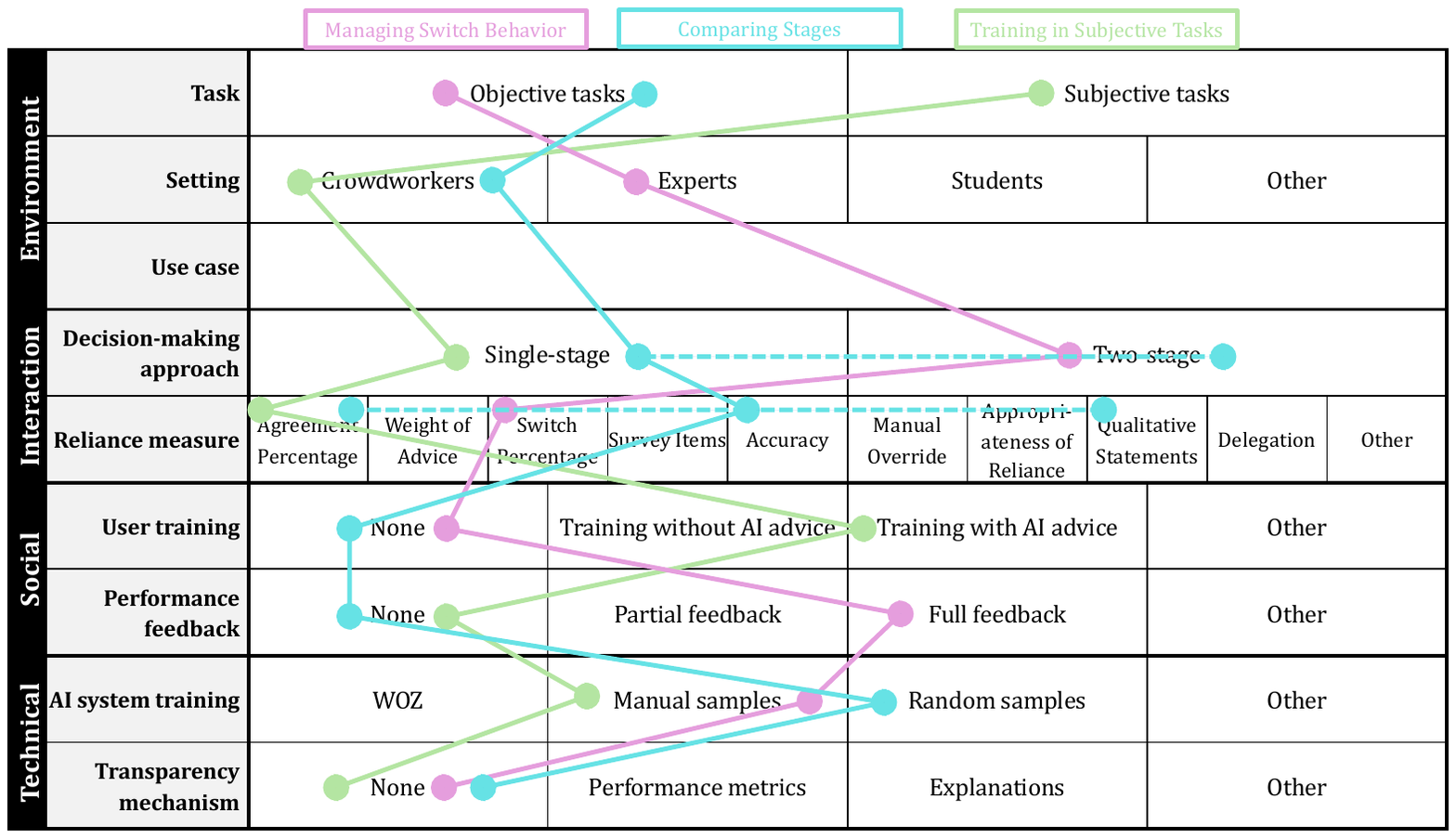}
    \caption{Prospective application of the morphological box on the three research streams: \textit{Managing Switch Behavior}, \textit{Comparing Stages}, and \textit{ Training in Subjective Tasks}.}
    \Description{The figure illustrates the application of the morphological box on the three research streams: \textit{Managing Switch Behavior}, \textit{Comparing Stages}, and\textit{ Training in Subjective Tasks}.}
    \label{fig:morph_boxapplicationfuture}
\end{figure}
\newpage

\section{Additional Tables}

\begin{table}[htbp]
    \centering
    \begin{tabularx}{\textwidth}{m{2cm}m{2cm}lm{1.5cm}X} \toprule
        \textbf{Article} & \textbf{Focus} & \textbf{Time Span} & \textbf{Method} & \textbf{Difference to this survey}  \\ \midrule
        \citet{adadi2018peekin} & Explainable AI & 2004-2018 & Survey & This article reviews existing approaches to achieving explainable AI, but does not consider human-AI interaction. \\ \hline
        \citet{rudresh2023explainable} & Explainable AI & until 2022 & Survey & The article considers programming techniques for explainable AI, but these only shed light on one part of the STS--the technical component.  \\ \hline
        \citet{mehrabi2021survey} & AI fairness & until 2021 & Survey & The article focuses on defining AI fairness as the output of the AI system and does not consider human-AI interaction. \\ \hline
        \citet{pessach2022fairness} & AI fairness & 2019 - 2021 & Survey & The article focuses on the main concepts of algorithmic fairness rather than highlighting the human–AI interaction. \\ \hline
        \citet{kaur2022trustworthy} & Trustworthy AI & until 2021 & Survey & The article focuses on trust and how it can be achieved, but does not consider the issue of reliance on AI. \\ \hline
        \citet{lai2023towards} & Human-AI Decision Making & 2018-2021 & Survey & The article addresses AI reliance, but only in a very general way as one aspect of human-AI interaction. There is a lack of insight into empirical research on AI reliance. \\ \hline
        \citet{guo2024decision} & AI Reliance & N/A & Conceptual & The article is purely conceptual, with no empirical data. \\ \bottomrule
    \end{tabularx}
    \caption{Related articles to this survey. Not all articles state the time span of their literature queries. For these articles, we assumed that the date of the last revision was the cut-off point.}
    \label{tab:related_work}
\end{table}

\begin{table}[htbp]
\def\arraystretch{1.5}%
    \centering
    \begin{tabularx}{\textwidth}{|m{3cm}|X|} \hline
        \textbf{Measure} & \textbf{Formula}  \\ \hline
        \textit{Agreement percentage (AP)} &  \textit{AP = \% of cases, where the AI advice and user decision coincide}  \\ \hline
        \textit{Weight of advice (WOA)} & $WOA = \frac{\text{final estimate - initial estimate}}{\text{advice - initial estimate}}$ \\ \hline
        \textit{Switch percentage (SP)} & \textit{SP = \% of cases, where the AI advice and user decision coincide} \\ \hline
        \textit{Accuracy (Acc)} & $Acc = \frac{\text{No. of decision instance with correct decision}}{\text{No. of all decision instances}} * 100\%$ \\ \hline
        \textit{Manual Override (MO)} & \textit{MO = \% of cases, where the user overrides the AI system's decision} \\ \hline
        \textit{Appropriateness of} \newline \textit{Reliance (AoR)} & 
        \textit{AoR = (Relative Self Reliance (RSR), Relative AI Reliance (RAIR))}, where \newline {\footnotesize RSR = \% cases users rejecting incorrect AI advice and remaining correct initial decisions, and} \newline {\footnotesize  RAIR =  \% cases users accepting correct AI advice and switching from incorrect initial decision} \\ \hline
        \textit{Delegation (Del)} & \textit{Del = \% of cases, where the user delegated the task to the AI system} \\ \hline
    \end{tabularx}
    \caption{Table of quantitative metrics}
    \label{tab:measures}
\end{table}

\end{document}